\documentclass[manuscript]{acmart}

\usepackage{subfigure}
\usepackage{hyperref}
\usepackage{amsfonts}
\usepackage{enumitem}

\AtBeginDocument{%
  \providecommand\BibTeX{{%
    \normalfont B\kern-0.5em{\scshape i\kern-0.25em b}\kern-0.8em\TeX}}}

\setcopyright{acmcopyright}
\copyrightyear{2021}
\acmYear{2021}
\acmDOI{10.1145/1122445.1122456}

\acmConference[RecSys '21]{RecSys '21: ACM Conference on Recommender Systems}{27th September-1st October, 2021}{Amsterdam, Netherlands}
\acmBooktitle{RecSys '21,
  27th September-1st October, 2021, Amsterdam, Netherlands}
\acmPrice{15.00}
\acmISBN{978-1-4503-XXXX-X/18/06}

\begin{document}

\title{Denoising User-aware Memory Network for Recommendation}

\author{Zhi Bian}
\authornote{Both authors contributed equally to this research.}
\email{bianzhi.bz@alibaba-inc.com}
\author{Shaojun Zhou}
\authornotemark[1]
\email{zsj148798@alibaba-inc.com}
\author{Hao Fu}
\email{hfu@mail.ustc.edu.cn}
\author{Qihong Yang}
\email{xiaokui.yqh@alibaba-inc.com}
\author{Zhenqi Sun}
\email{sunzhenqi.szq@alibaba-inc.com}
\author{Junjie Tang}
\authornote{Corresponding author.}
\email{lixi.tjj@alibaba-inc.com}
\affiliation{%
	\institution{Alibaba Group}
}
\author{Guiquan Liu}
\email{gqliu@ustc.edu.cn}
\affiliation{%
	\institution{University of Science and Technology of China}
}
\author{Kaikui Liu}
\email{damon@alibaba-inc.com}
\author{Xiaolong Li}
\email{xl.li@antfin.com}
\affiliation{%
  \institution{Alibaba Group}
}

\renewcommand{\shortauthors}{Zhi Bian and Shaojun Zhou, et al.}

\begin{abstract}
For better user satisfaction and business effectiveness, 
more and more attention has been paid to the sequence-based recommendation system, 
which is used to infer the evolution of users' dynamic preferences, 
and recent studies 
have noticed that the evolution of users' preferences can be better understood from the implicit and explicit feedback sequences. 
However, most of the existing recommendation techniques do not consider the noise contained in implicit feedback, 
which will lead to the biased representation of user interest and a suboptimal recommendation performance. 
Meanwhile, the existing methods utilize item sequence for capturing the evolution of user interest. 
The performance of these methods is limited by the length of the sequence, and can not effectively model the long-term interest in a long period of time. Based on this observation, we propose a novel CTR model named 
denoising user-aware memory network (DUMN). 
Specifically, the framework: (i) proposes a feature purification module based on orthogonal mapping, which use the representation of explicit feedback to purify the representation of implicit feedback, 
and effectively denoise the implicit feedback; 
(ii) designs a user memory network to model the long-term interests in a fine-grained way by improving the memory network, which is ignored by the existing methods; and (iii) develops a preference-aware interactive representation component to fuse the long-term and short-term interests of users based on gating to understand the evolution of unbiased preferences of users. Extensive experiments on two real e-commerce user behavior datasets show that DUMN has a significant improvement over the state-of-the-art baselines. 
The code of DUMN model has been uploaded as an additional material.

\end{abstract}




\maketitle

\section{Introduction}

Large-scale e-commerce platforms such as Taobao and Amazon have hundreds of millions of users' interaction data every day. Click-Through Rate (CTR) prediction plays an important role in the personalized recommendation system \cite{DBLP:journals/corr/HidasiKBT15,ni2018perceive,zhou2019deep,zhou2018deep}. It can recommend items consistent with users' interests and preferences by analyzing users' historical behavior data, thus greatly improving users' satisfaction and reducing information overload.


The key to CTR prediction is to understand the evolution of user preferences 
through historical behavior data. 
Traditional CTR prediction methods such as matrix factorization (MF) \cite{koren2009matrix} and collaborative filtering (CF) \cite{goldberg1992using} try to learn low-order cross features and capture the similarity between users and items through the user-item rating matrix constructed from user feedback data. 
With the rapid development of deep learning, CTR prediction method based on deep learning such as DeepFM \cite{guo2017deepfm} and Deep\&Cross \cite{wang2017deep} and xDeepFM \cite{lian2018xdeepfm} can effectively capture the high-order cross features of users and items, and capture the evolution of users' interests through LSTM/GRU network modeling click sequence \cite{li2017neural,kang2018self,tang2018personalized}. Methods such as DIEN \cite{zhou2019deep} and DSIN \cite{DBLP:conf/ijcai/FengLSWSZY19}, regard the user's click feedback data as sequence signals, and use RNN-based networks to summarize the user's preference. 
Recently, researchers point out that the modeling of click sequence can only focus on what users are interested in, but ignore the modeling of what users are not interested in, which leads to the captured user preferences are biased \cite{xie2020deep}. 
In view of this, the interaction data is subdivided into implicit and explicit feedback~\cite{gu2020deep,xie2020deep}. 
Among them, explicit feedback is defined as precise but relatively rare feedback that can directly indicate users' positive/negative preferences in the view page, such as rating and tagging like/dislike, while implicit feedback refers to rich but noisy feedback that contains noise and cannot directly indicate the user's preferences. Implicit feedback includes click and unclick. Click feedback indicates that the user clicks an item on the view page, while unclick feedback indicates that the user slides down but does not click. 
In general, click feedback may come from users accidentally clicking some wrong items; unclick feedback also includes items that the user may be interested in, but it scrolls too fast to notice. 
\cite{liu2010unifying} and \cite{hadash2018rank} model users' explicit negative feedback information through collaborative filtering (CF) and multi-task learning, which improves the performance of the model to a certain extent. \cite{lv2020unclicked}, \cite{xie2020deep} and \cite{zhao2018recommendations} consider modeling the unclicked sequence in the user's implicit negative feedback information.

Despite these methods have achieved significant performance by modeling both implicit and explicit feedback to understand the evolution of users' unbiased preferences, 
we argue that the inherent noises of implicit feedback are not dealt with effectively. 
The existing noise purification methods use attention mechanism \cite{xie2020deep} and autoencoders \cite{shenbin2020recvae} to increase the attention of related items, 
but the attention value itself may be inaccurate, which leads to room for improvement in the purification of noise features. 
Compared with the implicit feedback with noise, the explicit feedback can accurately indicate the user's preference. Therefore, 
the noise in the implicit feedback representation can be purified by the explicit feedback representation, 
and the user's preference can be more clearly described by the purified implicit representation and explicit representation, which has not been considered by existing methods.

In addition to the lack of noise purification, the existing methods are insufficient to represent the long-term preferences of users. 
Long-term preference refers to the behavior preference of users over a long period of time, which is usually relatively stable. 
The sequential recommendation methods, such as SDM \cite{lv2019sdm}, DIEN \cite{zhou2019deep} and DSIN \cite{DBLP:conf/ijcai/FengLSWSZY19}, try to increase the length of users' click sequence to capture users' stable long-term preferences. 
However, they have the following problems. 
First, 
the existing models of capturing long-term interest are all item-based methods, and the length of the item sequence used in these methods is often limited by the memory and computing resources, which leads to the gap between the model preference and the user's stable long-term preference. 
On the contrary, we argue that users' long-term interests should be characterized from the user level. Specifically, long-term interests should be sequence-independent and mainly related to the user profile. For example, people who own a pet cat may purchase certain cat food at intervals, which may not be related to their recent behavior sequences. Furthermore, the user's long-term interests characterized from a user-based perspective enables sequence decoupling, so as to ensure that the model is not limited to the length of the sequence. 
Second, the feedback information such as rating, click/unclick, like/dislike can reflect users' preference, and the preferences formed by different types of feedback are distinctive. Only by fine-grained modeling multiple types of sequential feedback can we better understand the long-term preferences of the user. 
MIMN \cite{pi2019practice} 
uses NTM \cite{graves2014neural} to maintain the latest interest state for each user and its update depends on the real-time user's click behavior to trigger events. 
However, all the methods mentioned above only use the click sequence to describe users' preferences, which leads to the incomplete description of user preferences. 
Based on the above analysis,  it can be seen that user's long-term preference modeling and noise feature purification are both unsolved problems in CTR prediction, and without loss of generality, we define the preference representation learned from the limited length sequence as short-term interest.

In seeking to address these challenges, we propose DUMN for recommendation. First, 
we design a feature purification (FP) module based on vector orthogonal mapping~\cite{qin2020feature} for short-term preference modeling. Specifically, we construct two sets of contrast pairs $<click,$ $dislike>$ and $<unclick, like>$, and map the first representation vector $click$ and $unclick$ of each pair to the vertical direction of the second representation vector $dislike$ and $like$, and the mapped vector is used as the purified representation. 
Then, a user memory network (UMN) is proposed to understand the stable long-term preference of users in a fine-grained way. UMN improves the memory network used in NTM \cite{graves2014neural} to represent all types of feedback, and designs a novel triple loss to regularize the memory network, so that 
the content written in the memory network can truly express the user's preferences. 
Finally, a preference-aware interactive representation (PAIR) module is proposed to obtain the cross representation that can simultaneously aggregate user long-term and short-term preferences. 
Our contributions are summarized as follows:
\begin{itemize}[leftmargin = 10 pt]
	\item We introduce a novel denoising user-aware memory network for CTR prediction task, which can understand the unbiased evolution of users' preferences by fine-grained modeling of users' feedback information. 
	\item We design a FP module for implicit feature purification. Through a novel vector orthogonal mapping method, the FP module can effectively extract the differences in two sets of contrast pairs $<click, dislike>$ and $<unclick, like>$.
	\item 
	We propose UMN module and PAIR module for fine-grained long-term preference modeling and cross representation of long-term and short-term interests, respectively.
	\item We conduct experiments on two real-world datasets. The experimental results show that our DUMN is superior to the existing state-of-the-art baselines, which verifies the effectiveness of our DUMN.
\end{itemize}



\section{Related Work}
\subsection{Recommendation with Implicit Feedback}
User feedback data contains both precise but relatively rare explicit feedback and rich but noisy implicit feedback. As mentioned in Introduction, 
we can obtain the user's unbiased preference presentation by modeling both explicit feedback and implicit feedback at the same time. 
A large number of existing works have also proved the improvement of system performance by using various feedback information such as implicit and explicit through experiments. 
\cite{hu2008collaborative} treated all unobserved items of users as negative instances and expressed the value of implicit feedback information as confidential. \cite{koren2008factorization,liu2010unifying,zhang2018coupledcf,shenbin2020recvae} fused CF and various feedback information, and \cite{jadidinejad2019unifying} used weak supervision method to model feedback information. Furthermore, more algorithms use multi-task learning frameworks to jointly solve ranking and rating tasks by combining various explicit feedback and implicit feedback \cite{hadash2018rank}. Recently, GCN-based GRCN \cite{wei2020graph} is proposed to implement pruning of noisy edges in the graph constructed by users' feedback information on items. FAWMF \cite{chen2020fast} applies variational autoencoder to realize adaptive weight assignment of implicit feedback and effective model learning. To the best of our knowledge, we are the first to use implicit feedback to purify the representation of 
feedback by orthogonal mapping.

\subsection{CTR Model}
Early methods of CTR prediction construct the interactive data of users and items into a user-item rating matrix, and the user-based or item-based CF method \cite{goldberg1992using} is used for rating prediction. In these methods, the user and item evaluation vectors are regarded as the presentation vectors of each user and item. Later, based on the co-occurrence matrix of CF algorithm, MF adds the concept of hidden vector to reduce the representation dimension of vector and enhance the ability of the model to deal with sparse matrix. MF based methods, such as singular value decomposition \cite{koren2009matrix}, non-negative matrix factorization \cite{fevotte2011algorithms} and probabilistic matrix factorization [32], are widely used in recommendation system.

%
%
%
%

With the rapid development of deep learning in many fields, such as computer vision \cite{huang2017densely,redmon2016you}, natural language processing \cite{devlin2019bert}, there are more and more researches on personalized recommendation of jobs \cite{borisyuk2017lijar}, music \cite{van2013deep}, news \cite{2014Personalized} and video \cite{covington2016deep}  based on deep neural network. Different from the traditional CTR prediction models such as MF and CF, which capture the similarity between users and items through feature engineering, the CTR prediction method based on deep learning uses neural networks to capture the interaction between features, which can better capture high-order interactive features~\cite{lian2018xdeepfm,li2020interpretable}.  Wide\&Deep \cite{cheng2016wide} considers the wide part for memory and the deep part for generalization together, making the model have the advantages of both logistic regression and deep neural network. DIN \cite{zhou2018deep} proposes an attention mechanism to capture the relative interest of candidates and obtain adaptive interest representation. 
Inspired by the success of the Transformer network in neural machine translation \cite{vaswani2017attention}, ATRANK \cite{zhou2018atrank} invented an attention-based framework for CTR prediction. DIEN \cite{zhou2019deep} proposed a two-layer RNN structure with an attention mechanism, which uses attention weights to control the second-layer RNN to activate the interests most relevant to the candidates. Furthermore, DSIN \cite{DBLP:conf/ijcai/FengLSWSZY19} captures the evolution of user preferences in the session by dividing the user's interactive behavior by session.

In recent years, inspired by the development of neural network graph embedding algorithm \cite{kipf2016semi}, more and more attention has been paid to graph structure development for various recommended scenarios \cite{jin2020efficient,shi2018heterogeneous}. \cite{yu2014personalized} use different types of entity relationships in heterogeneous information networks to make use of the personalized recommendation framework. 
Recently, in order to address the early summarization issue on heterogeneous information networks, NIRec \cite{jin2020efficient} is designed to capture and aggregate rich interactive patterns in both node-level and path-level. 

\section{Method}
\begin{figure*}
	\centering
	\includegraphics[width=\textwidth]{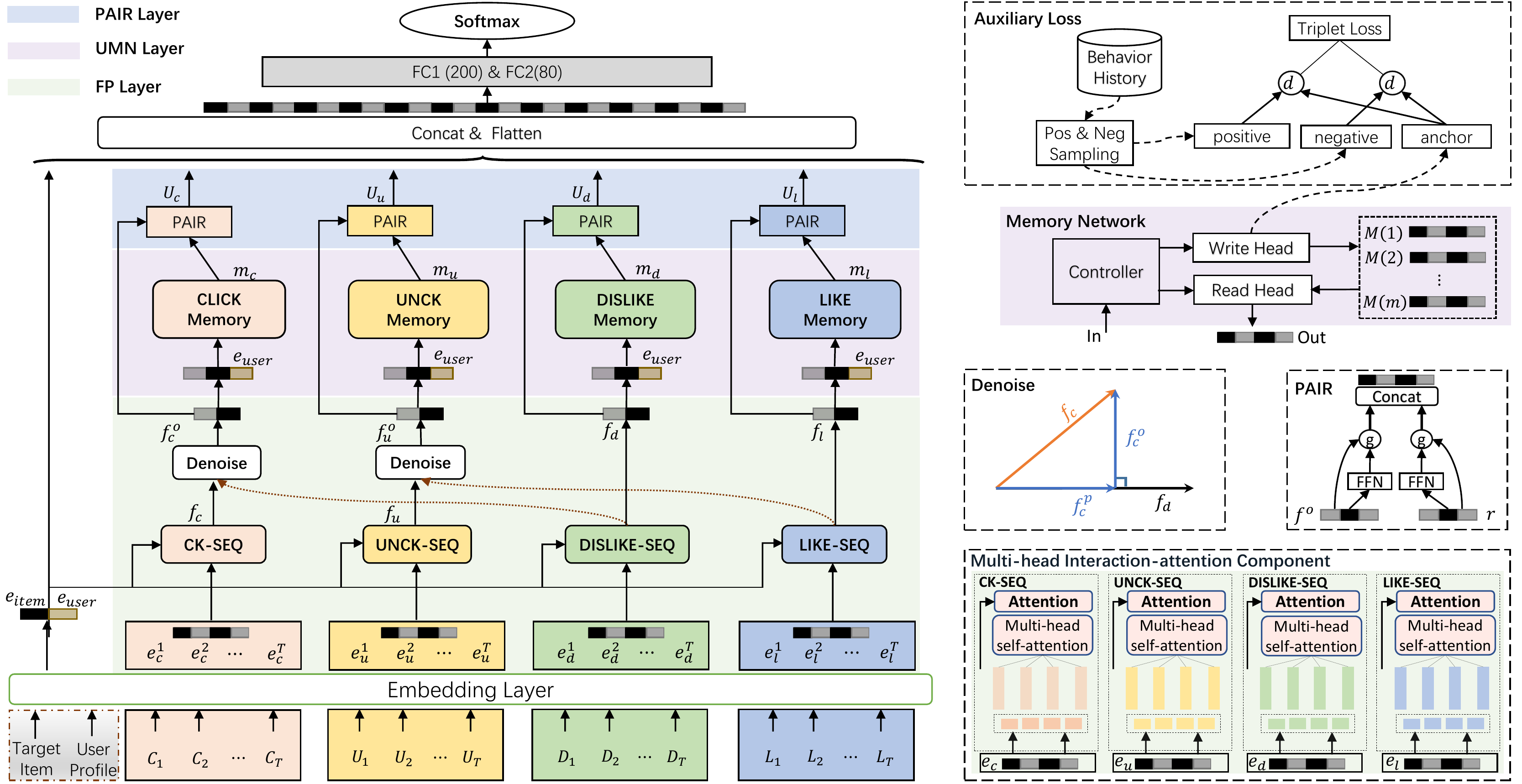}
	\caption{The overall architecture of denoising user-aware memory network.} \label{fig1}
\end{figure*}

In this section, we present the technical details of DUMN. As shown in Fig. \ref{fig1}, DUMN mainly consists of four modules, namely embedding layer, FP layer, UMN layer and PAIR layer. First, the embedding layer takes $user$, $ad$, user's $click$, $unclick$, $like$ and $dislike$ sequences as input, and implements embedding for all input data. 
Secondly, FP uses the multi-head interaction-attention component shown in the lower right side of Fig. \ref{fig1} to model the user's various implicit/explicit feedback sequences, and uses the representation of explicit feedback to purify the representation of implicit feedback by orthogonal mapping. 
Third, the UMN module captures users' fine-grained long-term interests by improving the memory network to all types of feedback. Finally, PAIR combines the short-term and the long-term interest representation to obtain the cross representation, and then uses the fully connected layer for CTR prediction.

\subsection{Problem Definition}

The DUMN network is represented by the function $\mathcal{F}(x;\theta)$, where $\theta$ is the parameters and $x$ is the input, which contains the initial representation of the user, target item and four sequences of user's click, unclick, like and dislike feedback. 
Our goal is to make the predicted value $\hat{y}$ of $\mathcal{F}(x;\theta)$ as close to the user's real click preference $y\in \{0,1\}$ as possible by minimizing the specific loss $\mathcal{L}$, so as to realize the purpose of CTR prediction.

\subsection{Embedding Layers}
The input of DUMN can be divided into three parts: $user\ profile$, $ad$ and $user\ behavior\ sequences$. The attribute fields of $user\ profile$ includes user\_id, gender, age, etc. $ad$ refers to the target item for CTR prediction, which includes item\_id and brand\_id. $user\ behavior\ sequence$ is a list of items. In this paper, we model four kinds of behavior sequences, which are click sequence $\textbf{C}=[C_1, C_2,...,C_T]$ and unclick sequence $\textbf{U}=[U_1, U_2,...,U_T]$ of implicit feedback, dislike sequence $\textbf{D}=[D_1, D_2,...,D_T]$ and like sequence $\textbf{L}=[L_1, L_2,...,L_T]$ of explicit feedback, where $T$ is the maximum sequence length input to the model. As previously did \cite{zhou2019deep,DBLP:conf/ijcai/FengLSWSZY19}, we use the embedding layer to transform the high dimensional sparse ids into low dimensional dense representations. The concatenate of different fields’ embedding output from $user\ profile$ and $ad$ form $e_{user}$ and $e_{item}$ respectively. Accordingly, after embedding $\textbf{C}$, $\textbf{U}$, $\textbf{D}$ and $\textbf{L}$, the outputs are $\textbf{e}_c\in{\mathbb{R}^{T\times E}}$, $\textbf{e}_u\in{\mathbb{R}^{T\times E}}$, $\textbf{e}_d\in{\mathbb{R}^{T\times E}}$ and $\textbf{e}_l\in{\mathbb{R}^{T\times E}}$, where $E$ is the unified dimension of embedding.




\subsection{Feature Purification Layer}
The feature purification layer takes the embedding lists $\textbf{e}_c$ and $\textbf{e}_u$ of implicit feedback, $\textbf{e}_l$ and $\textbf{e}_d$ of explicit feedback, $\textbf{e}_{user}$ of the user and $\textbf{e}_{item}$ of the target item as inputs. Specially, the feature purification layer 
uses two components to learn the short-term interest expressed by the feedback sequences and purify the feature of implicit feedback by proposing a novel vector orthogonal mapping method.

\subsubsection{Multi-head Interaction-attention Component}
Inspired by the potential of self-attention mechanism in data correlation learning \cite{vaswani2017attention}, we model various feedback behaviors of users on the framework of multi-head self-attention network to obtain fine-grained preference representation, 
and the whole process is shown in the lower right side of Fig. \ref{fig1}. We use $\textbf{e}_c$ as an example to introduce the working of multi-head self-attention network. Mathematically, 
we construct $\textbf{e}_c$ as a form with $H$ heads, that is $\textbf{e}_c = [\textbf{e}_{c,1},...,\textbf{e}_{c,h},...,\textbf{e}_{c,H}]$, 
where $\textbf{e}_{c,h}\in{\mathbb{R}^{T\times \frac{1}{H}E}}$ is the $h$-th head of $\textbf{e}_c$, $H$ is the number of heads. The output of the multi-head self-attention network is 
calculated as follows:


\begin{equation}\label{eq1}
	\textbf{head}_h={\rm softmax}(\frac{\textbf{e}_{c,h}\textbf{W}_{c,h}^Q(\textbf{e}_{c,h}\textbf{W}_{c,h}^K)^{\rm T}}{\sqrt{T}})\textbf{e}_{c,h}\textbf{W}_{c,h}^V\ \ \ h=1,2,...,H,
\end{equation}
\begin{equation}
	\textbf{O}_{c}={\rm Concat}(\textbf{head}_1,\textbf{head}_2,...,\textbf{head}_H)\textbf{W}^F
\end{equation}
where $\textbf{W}_{c,h}^Q$, $\textbf{W}_{c,h}^K$ and $\textbf{W}_{c,h}^V$ are trainable linear matrices. 
Then, we calculate the attention value between the target $ad$ item and the item representation of each position in the sequence representation through the full connection layer, as shown in the following formulas:
\begin{gather}
\alpha_j = {\rm ReLU}({\rm Concat}(\textbf{e}_{user},\textbf{e}_{item},\textbf{o}_{c}^j)\textbf{W}_{c})\ \ \ j=1,2,...,T,\\
\tilde{\alpha}_j = \frac{{\rm exp}(\alpha_j)}{\sum_{j'=1}^{T}{\rm exp}(\alpha_{j'})}\ \ \ j=1,2,...,T,
\end{gather}
where $\textbf{o}_{c}^j$ is the output representation of the $j$-th item through the multi-head self-attention network. $\textbf{W}_{c}$ is trainable linear matrix. ${\rm ReLU}(\cdot)$ is the activation function. 
Then, the representation of the click sequence fused with the target item can be calculated as:
\begin{equation}\label{eq5}
\textbf{f}_{c} = \sum_{j=1}^{T}\tilde{\alpha}_j\textbf{o}_{c}^j	.
\end{equation}

Similarly, we can get the representations of unclick, like, and dislike sequences as $\textbf{f}_{u}$, $\textbf{f}_{l}$, and $\textbf{f}_{d}$ by using formulas \eqref{eq1}-\eqref{eq5}.

\subsubsection{Feature Orthogonal Mapping Component}

The representations $\textbf{f}_{c}$ and $\textbf{f}_{u}$ of the implicit feedback contain inherent noise, and the goal of the feature orthogonal mapping component is to purify the representations $\textbf{f}_{c}$ and $\textbf{f}_{u}$ of implicit feedback. 
We argue that the explicit feedback representation which can definitely indicate user preferences can be used to purify the implicit feedback representation which can not directly indicate user preferences. So we extract two groups of orthogonal mapping pairs $<click,dislike>$ and $<unclick,like>$ with differences from implicit and explicit feedback, and their corresponding sequence representations are $<\textbf{f}_{c},\textbf{f}_{d}>$ and $<\textbf{f}_{u},\textbf{f}_{l}>$ respectively. Taking $<\textbf{f}_{c},\textbf{f}_{d}>$ as an example, in order to purify the representation of implicit feedback sequence with noise, we project the first element $\textbf{f}_{c}$ of the sequence representation pair onto the orthogonal direction of the second element $\textbf{f}_{d}$. The original feature vector $\textbf{f}_{c}$ is projected into the orthogonal feature space to eliminate the noise features. 
Compared with the original vector, the orthogonal mapping vector contains pure and efficient user preferences.

Formally, we describe the orthogonal mapping process of sequence pair $<\textbf{f}_{c},\textbf{f}_{d}>$ in a two-dimensional space shown in the middle of the right side of Fig. \ref{fig1}.  The noise representation vector $\textbf{f}_{c}^p$ in $\textbf{f}_{c}$ can be obtained by projecting the vector $\textbf{f}_{c}$ onto the vector $\textbf{f}_{d}$:
\begin{equation}\label{eq6}
\textbf{f}_{c}^p={\rm project}(\textbf{f}_{c},\textbf{f}_{d}),
\end{equation}
where ${\rm project}(a,b)=\frac{a\cdot b}{|b|}\frac{b}{|b|}$ represent the projection function. $a$ and $b$ are vectors with the same dimension. Then, we can get the vector representation purified by orthogonal mapping as follows:
\begin{equation}\label{eq7}
\textbf{f}_{c}^o=\textbf{f}_{c}-\textbf{f}_{c}^p.
\end{equation}

Obviously, according to formula \eqref{eq6}, the representation of implicit feedback contains a mixture of user's click and dislike noise $\textbf{f}_{c}^p$, which is filtered from the original representation $\textbf{f}_{c}$, and the new pure feature $\textbf{f}_{c}^o$ can 
effectively represent the user's pure click preference in orthogonal space, 
which is also in line with the assumption that the user's click and dislike representation should be distinctive. Similarly, according to formulas \eqref {eq6}-\eqref{eq7}, we can get the orthogonal mapping vector $\textbf{f}_{u}^o$ between the representation $\textbf{f}_{u}$ of unclick sequence and the representation $\textbf{f}_{l}$ of like sequence, 
and the purified vectors $\textbf{f}_{c}^o$ and $\textbf{f}_{u}^o$ can better describe the unbiased preferences of users.


\subsection{User Memory Network Layer}

In order to get a more stable and fine-grained representation of long-term preference from the perspective of users than the item-based methods, 
we improve the memory network used in NTM  \cite{graves2014neural}. 
Specifically, the memory network of NTM contains multiple slots to model the user's click sequence, and uses a controller to generate the key for reading or writing of the user's click sequence representation, so as to complete the operation of \textbf{memory read} and \textbf{memory write} for the memory network. Considering that memory network can store feature representation, and each slot has the characteristic of aggregating the same feature representation, we extend it to store user-level long-term interests, 
so that the user feature representation in the same slot can reflect the similar interests between users, and the long-term interest representation obtained in this way is more generalized than using only user\_id embedding.

We improve the basic memory network as follows. First, in order to capture users' fine-grained unbiased long-term interests, we use four memory network $\textbf{M}_{c}$, $\textbf{M}_{u}$, $\textbf{M}_{l}$ and $\textbf{M}_{d}$ to save users' click, unclick, like and dislike preferences respectively, and each memory network contains $m$ slots whose output dimension is $Z$. 
Second, the input of the controller is replaced by the concatenate of users' short-term representation obtained by FP and the embedding of $user\ profile$ from the representation of item to ensure that the model can learn the user level long-term interest representation. 
Taking $\textbf{M}_{c}$ as an example, \textbf{memory read} and \textbf{memory write} of $\textbf{M}_{c}$ are as follows.

\textbf{Memory read.} Input the concatenate of $\textbf{f}_{c}^o$ and $\textbf{e}_{user}$, and the controller generates a read key $\textbf{k}_c$ to address the memory $\textbf{M}_c$ through a fully connected layer. 
\begin{equation}
\textbf{k}_c={\rm FFN}({\rm Concat}(\textbf{f}_{c}^o,\textbf{e}_{user})),
\end{equation}
where ${\rm FFN}(\cdot)$ denotes the feed-forward network. Then, by traversing all memory slots, a weight vector $\textbf{w}_c^r$ is generated:
\begin{gather}
\textbf{w}_c^r(j)=\frac{{\rm exp}({\rm K}(\textbf{k}_c,\textbf{M}_c(j)))}{\sum_{j'=1}^{m}{\rm exp}({\rm K}(\textbf{k}_c,\textbf{M}_c(j')))}\ \ j=1,2,...,m,\label{eq9}\\
{\rm K}(\textbf{k}_c,\textbf{M}_c(j))=\frac{\textbf{k}_c^{\rm T}\textbf{M}_c(j)}{\left \|\textbf{k}_c \right \|\left \|\textbf{M}_c(j) \right \|},
\end{gather}
finally, the weighted memory summary is calculated as the output $\textbf{r}_c\in\mathbb{R}^Z$:
\begin{gather}
\textbf{r}_c=\sum_{j=1}^{m}\textbf{w}_c^r(j)\textbf{M}_c(j).
\end{gather}

\textbf{Memory write.} The generation of the weight vector $\textbf{w}_c^w$ for memory write is similar to the \textbf{memory read} operation in equation \eqref{eq9}. The controller also generates two additional keys, add vector $\textbf{add}_c$ and erase vector $\textbf{erase}_c$, to control the update of memory.
\begin{equation}
\textbf{M}_c=(1-\textbf{E}_c)\odot \textbf{M}_c+\textbf{A}_c,
\end{equation}
where $\textbf{E}_c=\textbf{w}_c^w\otimes \textbf{erase}_c$ is the erase matrix. $\textbf{A}_c=\textbf{w}_c^w\otimes \textbf{add}_c$ is the add matrix. $\odot$ and $\otimes$ means dot product and outer product respectively.

Accordingly, we can get four representations of users' long-term preferences: $\textbf{r}_{c}$, $\textbf{r}_{u}$, $\textbf{r}_{l}$ and $\textbf{r}_{d}$. 

\subsection{Preference-aware Interactive Representation Component}\label{section3_5}
In order to get the cross representation of users' long-term and short-term interests, we design a gating mechanism to fuse the long-term and short-term representations of users' preferences with the same type. Similarly, we use the representation of click preference as an example to get the cross representation $\textbf{U}_{c}$ of long-term and short-term click preference through the following formula:
\begin{equation}
\textbf{U}_c={\rm Concat}(\textbf{f}_{c}^o*{\rm sigmoid}(\textbf{f}_{c}^o\textbf{W}_1),\textbf{r}_{c}*{\rm sigmoid}(\textbf{r}_{c}\textbf{W}_2)),
\end{equation}
where $*$ is the Hadamard product. 
${\rm sigmoid(\cdot)}$ is the activation function and $\textbf{W}$ is the dimension conversion matrix to ensure that the dimensions of the two vectors are consistent. The cross representation $\textbf{U}_{u}$, $\textbf{U}_{l}$ and $\textbf{U}_{d}$ corresponding to unclick, like and dislike can also be obtained. Therefore, we get the deep cross representation interest representation of users:
\begin{equation}
\textbf{R}_{cross}={\rm Concat}(\textbf{U}_{c},\textbf{U}_{u},\textbf{U}_{l},\textbf{U}_{d}).
\end{equation}

Finally, we concatenate the representations of user, target item, long-term interests, short-term interests and cross features as deep representations, and then use a fully connected layer with ${\rm sigmoid}(\cdot)$ function to generate the predicted CTR as:
\begin{equation}
\hat{y}={\rm sigmoid}({\rm FFN}({\rm Concat}(\textbf{e}_{user},\textbf{e}_{item},\textbf{R}_{cross}))).
\end{equation}

\subsection{Loss Function}
In our proposed model DUMN, we have two goals: 1) the predicted CTR of the target item should be as close to the true label as possible; and 2) we need to ensure that the content written into these 4 memory networks can truly express the user's long-term preferences of click, unclick, like and dislike.

For goal 1), we achieve it by minimizing the average logistic loss as:
\begin{equation}
\mathcal{L}_1=-\frac{1}{N}\sum_{(x,y)\in \mathcal{D}}(y{\rm log}(\hat{y})+(1-y){\rm log}(1-\hat{y})),
\end{equation}
where $\mathcal{D}$ is the training set of size $N$. $x$ is the input of DUMN. $y\in \{0,1\}$ represents whether the user clicks the target item.

For goal 2), we propose an auxiliary loss based on triplet loss to help complete the \textbf{memory write} operation. Specifically, for each content update, we randomly sample an item from each of the four feedback sequences of click, unclick, like and dislike in each batch, 
and record their output $\textbf{s}_{c}$, $\textbf{s}_{u}$, $\textbf{s}_{l}$ and $\textbf{s}_{d}$ in the embedding layer. 
It is worth noting that the samples are sampled from the data of all user interactions, not only from the constructed sequence for short-term interest representation. Then, we construct positive and negative sample pairs $<\textbf{s}_{c},\textbf{s}_{u}>$, $<\textbf{s}_{u},\textbf{s}_{c}>$, $<\textbf{s}_{l},\textbf{s}_{d}>$ and $<\textbf{s}_{d},\textbf{s}_{l}>$ for $\textbf{M}_{c}$, $\textbf{M}_{u}$, $\textbf{M}_{l}$ and $\textbf{M}_{d}$ respectively. Without loss of generality, we use $\textbf{M}_{c}$ as an example to explain the proposed triplet loss:
\begin{equation}
\mathcal{L}_{c}={\rm max}({\rm d}(\textbf{q}_{c},\textbf{s}_{c})-{\rm d}(\textbf{q}_{c},\textbf{s}_{u})+margin,0),
\end{equation}
\begin{equation}
\textbf{q}_c=\sum_{j=1}^{m}\textbf{w}_c^w(j)\textbf{M}_c(j),
\end{equation}
\begin{equation}
{\rm d}(a,b)=1-\frac{a\cdot b}{\left \|a\right \|\left \|b\right \|},
\end{equation}
where ${\rm d}(\cdot,\cdot)$ is the function to calculate the similarity, and cosine similarity is used in this paper. $\textbf{s}_c$, $\textbf{s}_u$ and $\textbf{q}_c$ are positive sample, negative sample and anchor corresponding to the upper right side of Fig. \ref{fig1}, respectively. Similarly, we can get the triplet losses $\mathcal{L}_{u}$, $\mathcal{L}_{l}$ and $\mathcal{L}_{d}$ of $\textbf{M}_{u}$, $\textbf{M}_{l}$ and $\textbf{M}_{d}$, respectively. Furthermore, goal 2) can be achieved by accumulating these 4 auxiliary losses as:
\begin{equation}
\mathcal{L}_2=\mathcal{L}_c+\mathcal{L}_u+\mathcal{L}_l+\mathcal{L}_d.
\end{equation}
then, the final loss function of DUMN is expressed as $\mathcal{L}=\mathcal{L}_1+\mathcal{L}_2$.

\section{Experiments}
In this section, we conduct experiments on different datasets to prove the effectiveness of our DUMN by comparing with several state-of-the-art methods. 
We start with 3 research questions (RQ) to guide the experiment and the following discussion:
\begin{itemize}[leftmargin = 10 pt]
	\item \textbf{RQ1: }Compared with state-of-the-art methods, can DUMN achieve better performance?
	\item \textbf{RQ2: }What is the impact of modules designed in DUMN? Are the proposed feature purification layer and user memory network layer modules necessary for improving performance?
	\item \textbf{RQ3: }What is the impact of hyper-parameter settings on CTR prediction performance in DUMN?
\end{itemize}

\subsection{Experimental Setups}
\subsubsection{Data Description}
We evaluate the model on two real-world e-commerce datasets, namely Alibaba dataset and Industrial dataset. For the Alibaba dataset, 
it is a public dataset, collecting ad display/feedback logs of 1.14 million users in Taobao's recommendation system. We have made statistics on the training set and test set of Alibaba dataset, including 26 million pieces of feedback records from 1.14 million users in 0.8 million items, and these items can be divided into 12,960 categories. 
For Industrial dataset, it contains 5.5 billion feedback records from 1.1 billion users and 91.0 million items in 30 days. Records from 2020-12-25 to 2021-01-18 are for training, and records from 2021-01-19 to 2021-01-24 are for testing. 
In particular, these two datasets contain a variety of implicit and explicit feedback data, we record the items purchased, added to the shopping cart and labeled with like in the log feedback as explicit like feedback, the items marked with dislike as explicit dislike feedback, the items simply clicked by user as implicit click feedback, the items displayed but not operated as implicit unclick feedback. 
Since we focus on the CTR prediction tasks, we treat all valid click interactions with the label of 1.

\subsubsection{Compared Methods}
We compared DUMN with following mainstream CTR prediction methods:

\begin{itemize}[leftmargin = 10 pt]
	\item \textbf{Wide\&deep: }Wide\&Deep \cite{cheng2016wide} consists of two parts: wide module of memory and deep module of generalization. It can effectively capture the high-order cross features between users and items.
	\item \textbf{PNN: }PNN \cite{qu2016product} uses the product layer containing inner product and outer product to capture the interactive patterns between interfield categories.
	\item \textbf{DeepFM: }DeepFM \cite{guo2017deepfm} is achieved by replacing the wide component in Wide\&Deep with FM layer, which can also capture the cross features of users and items.
	\begin{table}[!htp]
		\begin{minipage}[!t]{0.3\textwidth}
			\caption{Performance comparison with different baselines.}
			\label{table2}
			\centering
			\setlength{\tabcolsep}{1mm}{
				\begin{tabular}{ccc}
					\toprule
					Model&Alibaba&Industrial\\
					\midrule
					Wide\&Deep & 0.6326&0.7526  \\
					PNN & 0.6328&0.7602 \\
					DeepFM &0.6347 &0.7612  \\
					DIN & 0.6330& 0.7653 \\
					DIEN &0.6343 & 0.7803 \\
					DSIN & 0.6375& 0.7873 \\
					AutoInt & 0.6360& 0.7708 \\
					DFN & 0.6368& 0.7884\\
					DMT & 0.6378 & 0.8039 \\
					\midrule
					DUMN & \textbf{0.6496}& \textbf{0.8136}\\
					
					\bottomrule
				\end{tabular}
			}
		\end{minipage}
		%
		\begin{minipage}[!t]{0.33\textwidth}
			\caption{Effect of User Memory Network Layer.}
			\label{table4}
			\centering
			\setlength{\tabcolsep}{1mm}{
				\begin{tabular}{ccc}
					\toprule
					Model&Alibaba&Industrial\\
					\midrule
					DUMN-UMN & 0.6434&0.7966\\
					DUMN+UMN1 & 0.6446&0.8073\\
					DUMN & \textbf{0.6496}& \textbf{0.8136}\\
					\bottomrule
					
				\end{tabular}
			}
		\end{minipage}
		\begin{minipage}[!t]{0.35\textwidth}
			\caption{Effect of NTM.}
			\label{table4_}
			\centering
			\setlength{\tabcolsep}{1mm}{
				\begin{tabular}{ccc}
					\toprule
					Model&Alibaba&Industrial\\
					\midrule
					AutoInt&0.6360&0.7708\\
					AutoInt+UMN&0.6396&0.7802\\
					\midrule
					DFN&0.6368&0.7884\\
					DFN+UMN&0.6437&0.7973\\
					\bottomrule
				\end{tabular}
			}
		\end{minipage}
	\end{table}
	\item \textbf{DIN: }DIN \cite{zhou2018deep} uses attention mechanism to activate related users’ historical behaviors, which can fully exploits the relationship between users’ historical behaviors
	and the target item.
	\item \textbf{DIEN: }DIEN \cite{zhou2019deep} uses two layers of GRU to extract latent temporal interests from user behaviors and models interests evolving process.
	\item \textbf{DSIN: }DSIN \cite{DBLP:conf/ijcai/FengLSWSZY19} divides the user's historical click sequence into sessions, and then apply Bi-LSTM to model how users’ interests evolve and interact among sessions.
	\item \textbf{AutoInt: }AutoInt \cite{song2019autoint} introduces self-attentive neural network to find low-dimensional representations of the sparse and high-dimensional raw features and their meaningful combinations.
	\item \textbf{DMT: }DMT \cite{gu2020deep} uses multiple Transformers to model users’ multiple types of behavior sequences, including click sequence of items, cart sequence of items, and order sequence of items.
	\item \textbf{DFN: }DFN \cite{xie2020deep} jointly consider explicit and implicit feedback to learn user unbiased preferences for recommendation. Among them, the explicit feedback includes dislike sequence, and the implicit feedback includes click sequence and unclick sequence.
\end{itemize}

We divide the above baselines into the following two categories: the first is the non-sequence method that does not construct feedback sequence to obtain cross features between users and items, including Wide\&Deep, PNN and DeepFM; the second is the sequence-based method that uses feedback sequence information to capture users' interest evolution, including DIN, DIEN, DSIN, AutoInt, DMT and DFN. In particular, DMT and DFN are models based on implicit feedback and explicit feedback.
\subsubsection{Parameter Settings}
The DUMN model is implemented in Python based on the Tensorflow framework. All the experiments are conducted on a server machine equipped with a 16 GB NVIDIA Tesla V100 GPU. For hyper-parameters, the maximum length $T$ of each feedback sequence is set to 100. The output dimension of the embedding layer is 16. The dimension of the feed-forward network used in the memory network is set to 512. The number of the slot in the memory network is set to 256 and the dimension of each slot is set to 64. During the training phase,  we set the learning rate to 0.005 respectively, and use Adam as the optimizer. It is worth noting that the optimal parameters of DUMN are obtained by grid search, and the parameter sensitivity experiments of some important parameters are recorded in Section \ref{section4.4}. 
For the performance of Alibaba dataset, the baseline methods such as AutoInt, DMT and DFN are implemented per their GitHub settings, and the rest baseline methods use the performance recorded in \cite{DBLP:conf/ijcai/FengLSWSZY19}.

\subsection{Result Analysis (RQ1)}

We evaluated the click-through rate prediction (CTR) task in the datasets mentioned above. Specifically, the Area Under Curve (AUC) widely used in binary classification problems is used as the metric. 
The results of the evaluation on two datasets are recorded in Table \ref{table2}. 
By comparing with several state-of-the-art baselines, the following three conclusions can be drawn: 
\textbf{1)} Our proposed DUMN outperforms all baselines on these two datasets. The experimental results show that our proposed model is effective on the CTR prediction task, and the recommendation performance can be significantly improved by feature purification of implicit feedback representation and fusion of short-term and long-term interests. Compared with the best experimental results of baselines, the performances of DUMN on Alibaba dataset and Industrial dataset are improved by 1.18\% and 0.97\%, respectively. \textbf{2)} Among the two kinds of baselines, most sequence-based methods (DIN, DIEN, DSIN, AutoInt, DMT and DFN) are better than non-sequence methods (Wide\&Deep, PNN and DeepFM) in most cases, and an intuitive explanation is that these sequential methods can better capture the evolution of users' interests over time than non-sequential methods. 
It is worth noting that our DUMN also uses user feedback sequence, and uses attention mechanism to understand the evolution of user interest. 
\textbf{3)} In the sequence-based method, DFN achieves relatively good results, which is close to the experimental results of DSIN. Specifically, it can better get the unbiased preference evolution of users by describing the user feedback data in a more fine-grained way. 
It is also worth noting that our DUMN further divides user interests into long-term interests and short-term interests, which can obtain more accurate and deeper unbiased user preference representation.

\subsection{Ablation Study (RQ2)}
To investigate the effectiveness of components in DUMN, we conduct extensive ablation studies on the two datasets.


\subsubsection{Effect of Feature Purification}



The purpose of feature purification can purify the presentation of implicit feedback sequences with noise. 
In order to explore the effectiveness of using explicit feedback representation to denoise implicit feedback representation by orthogonal mapping in the FP component, we design DUMN-FP which removes the FP component for denoising. 
The evaluation results of DUMN-FP in Alibaba dataset and Industrial dataset are 0.6417 and 0.7981 respectively. Comparing the results with those of DUMN reported in Table \ref{table2}, we can find that: It is effective to purify the noise by using the proposed orthogonal mapping method. An intuitive explanation is that the explicit representation can accurately represent a user's preference, and the orthogonal space obtained by it should also be a noise-free space, which can better describe the user's unbiased preferences.

\subsubsection{Effect of User Memory Network Layer}

The user memory network layer is a fine-grained long-term interest characterization module based on user profile. In order to verify its effectiveness, we designed the following two models: DUMN-UMN which removes the user memory network layer and DUMN+UMN1 which 
replaces four memory networks with one memory network. 
Table \ref{table4} shows the evaluation results. It can be seen from the result that removing the representation of long-term interest or not distinguishing various feedback information will significantly reduce the prediction performance of the model, which is also in line with our previous hypothesis that only by fine-grained modeling multiple types of sequential feedback can we better understand the long-term preferences of the user.


Then, we further explore whether the capture of long-term interest can improve other existing models. Therefore, 
we introduce the user memory network which can capture users' long-term interest into AutoInt and DFN to get models AutoInt+UMN and DFN+UMN. Table \ref{table4_} shows the evaluation results, 
AutoInt+UMN and DFN+UMN beats AutoInt and 
\begin{table}[!htp]
	\begin{minipage}[!t]{0.33\textwidth}
		\caption{Effect of PAIR.}
		\label{table5}
		\centering
		\setlength{\tabcolsep}{1mm}{
			\begin{tabular}{ccc}
				\toprule
				Model&Alibaba&Industrial\\
				\midrule
				DUMN$_{CONCAT}$ & 0.6438&0.8023\\
				DUMN$_{CROSS}$ & 0.6487&0.8049\\
				DUMN$_{FFN}$ & 0.6486&0.8037\\
				DUMN$_{ATTE}$ & 0.6433&0.7966\\
				DUMN & \textbf{0.6496}& \textbf{0.8136}\\
				\bottomrule
			\end{tabular}
		}
	\end{minipage}
	\begin{minipage}[!t]{0.33\textwidth}
		\caption{Effect of the Proposed Triplet Loss.}
		\label{table6}
		\centering
		\setlength{\tabcolsep}{1mm}{
			\begin{tabular}{ccc}
				\toprule
				Model&Alibaba&Industrial\\
				\midrule
				DUMN-TL & 0.6467&0.7969\\
				DUMN$_{RS}$ & 0.6472&0.7991\\
				DUMN & \textbf{0.6496}& \textbf{0.8136}\\
				\bottomrule
			\end{tabular}
		}
	\end{minipage}
	\begin{minipage}[!t]{0.3\textwidth}
		\centering
		\makeatletter\def\@captype{table}\makeatother\caption{Effect of Implicit Feedback.}
		\label{table7}
		\begin{tabular}{ccc}
			\toprule
			Model&Alibaba&Industrial\\
			\midrule
			DUMN$_{IF}$ & 0.6392&0.8072\\
			DUMN$_{AF}$ & 0.6394&0.8091\\
			DUMN & \textbf{0.6496}& \textbf{0.8136}\\
			\bottomrule
		\end{tabular}
	\end{minipage}
\end{table}
DFN respectively, 
showing 
that the performance of existing methods has been improved after adding NUM module 
which can capture users' long-term interests, indicating that the integration of long-term and short-term interests can better understand the evolution of users' preferences.



\subsubsection{Effect of Preference-aware Interactive Representation Component}
The purpose of the preference-aware interactive representation component is to fuse short-term and long-term interests. In order to explore the impact of different fusion methods on CTR prediction performance, we design a variety of fusion methods and get the following four methods: DUMN$_{CONCAT}$ which use concatenate operation, DUMN$_{CROSS}$ which use cross operation, DUMN$_{FFN}$ which use the feed-forward network to represent first and then concatenate, and DUMN$_{ATTE}$ which use attention operation. Among them, concatenate operation represents the function ${\rm Concat}(\cdot)$, and cross operation refers to the concatenate after the 
addition, subtraction and multiplication of the elements at the corresponding positions of two vectors. According to Table \ref{table5}, the DUMN uses the gate-based method proposed in Section \ref{section3_5} can achieve the best performance on the fusion of long-term and short-term interest representation, which proves the feasibility of our proposed gate-based fusion method. In addition, we find that the cross and attention mechanism operations that can get high-order crossover features are better than the direct concatenate operation. 
The intuitive explanation is that the user's preference for the target item needs to be represented by both long-term interest and short-term interest, and the dependence of different items on the long-term and short-term is also different.


\subsubsection{Effect of the Proposed Triplet Loss}
The triplet loss is to ensure that the content written into the 4 memory networks can truly express the user's preferences. To evaluate the proposed triplet loss, we design a variety of comparison methods: DUMN-TL which removes the triplet loss from the loss function and DUMN$_{RS}$ which uses random sampling to sample positive and negative samples. It should be noted that our DUMN uses cosine similarity to find the hardest positive and negative samples in each batch. According to the results shown in Table \ref{table6}, we can find that the performance of our DUMN is significantly reduced by removing the limit of triple loss, and different sampling samples also have an impact on the performance. Specifically, 
the more dissimilar the representations of the sample pairs are, 
the more effective it is to ensure that the contents written in the memory network are reliable.

\subsubsection{Effect of Implicit/Explicit Feedback}

The intention of the fine-grained description of various implicit and explicit feedback is to obtain unbiased user preferences. We design two models to evaluate the improvement of CTR prediction brought by fine-grained interest representation: DUMN$_{IF}$ which uses only implicit feedback and removes $\textbf{M}_l$ and $\textbf{M}_d$ in user memory network layer, DUMN$_{AF}$ which connects all types of feedback into a sequence through time. It should be noted that these new models do not use orthogonal mapping to purify the representation of implicit feedback, and DUMN$_{AF}$ 
constructs a single sequence of all feedback information, 
which makes it difficult to capture the fine-grained interest representation formed by different types of user feedback. 
Table \ref{table7} shows the performance 
\begin{minipage}[ht]{\textwidth}
	\begin{minipage}[!t]{0.5\textwidth}
		\centering
		\begin{minipage}[!t]{\textwidth}
			\centering
			\includegraphics[width=.65\textwidth]{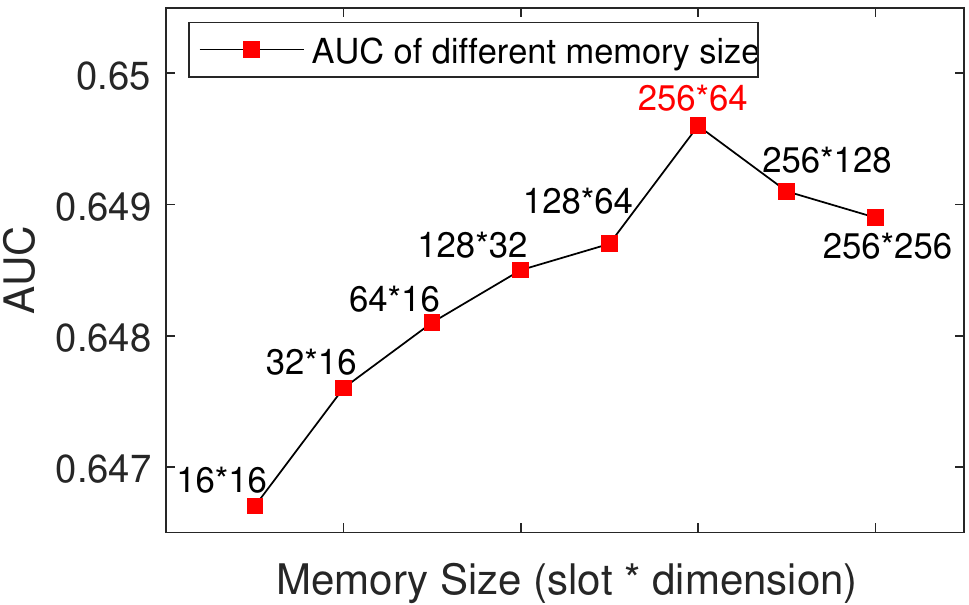}
			\makeatletter\def\@captype{figure}\makeatother\caption{Impact of \texorpdfstring{$m$}{} and \texorpdfstring{$Z$}{}.} \label{fig-4.4}

			\makeatletter\def\@captype{table}\makeatother\caption{Results of dislike prediction on Alibaba dataset.}
			\label{table4.5}
			\addvspace{0.3cm}
			\begin{tabular}{ccc}
				\toprule
				Model&Alibaba-dislike&Industrial-dislike\\
				\midrule
				DFN & 0.7609&0.7394\\
				DUMN & \textbf{0.8062}&\textbf{0.7601}\\
				\bottomrule
			\end{tabular}
		\end{minipage}
	\end{minipage}
	\begin{minipage}[ht]{0.47\textwidth}
		\centering
		\makeatletter\def\@captype{figure}\makeatother\subfigure[]{
			\begin{minipage}[t]{0.48\textwidth}
				\centering
				\includegraphics[width=\textwidth]{
					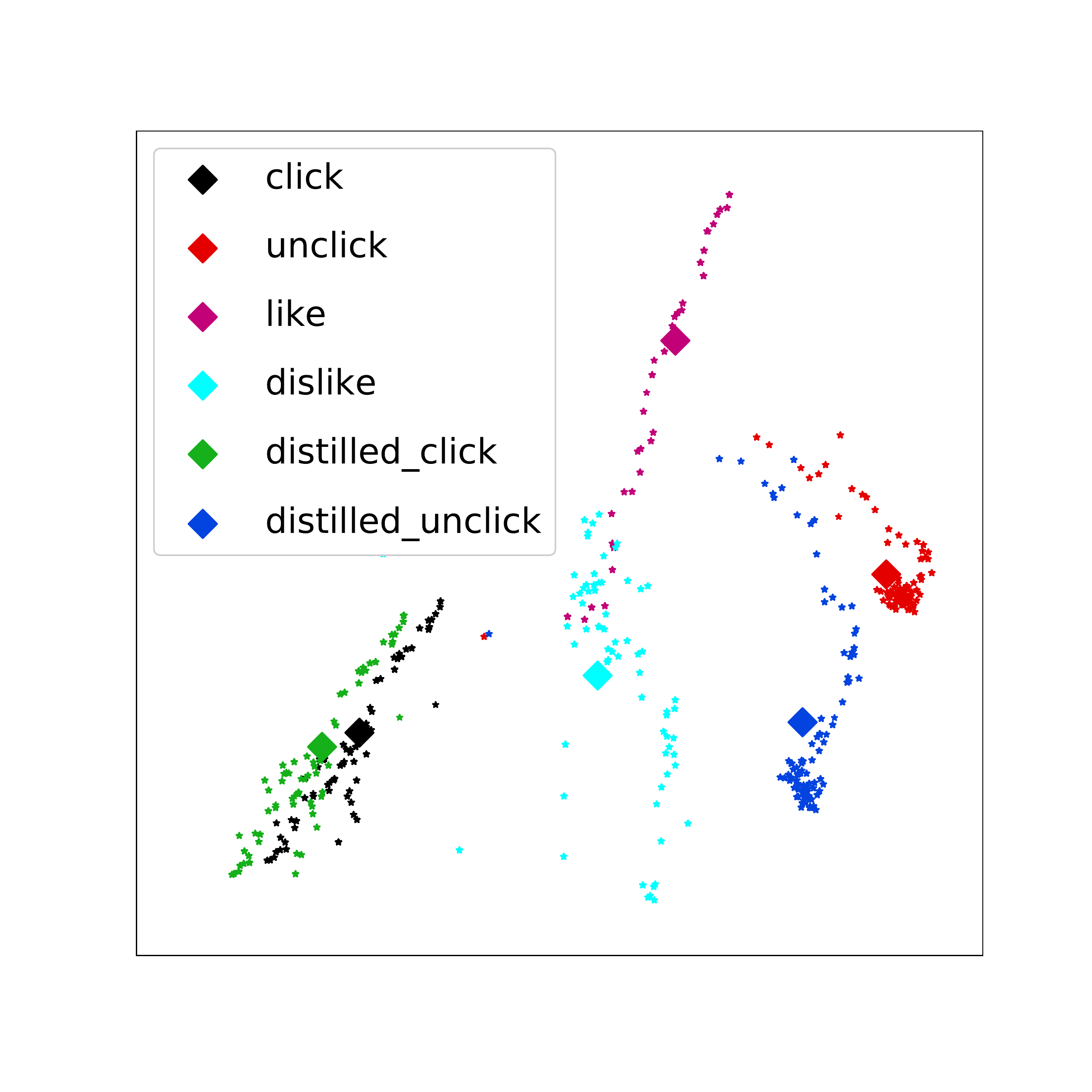}
				
			\end{minipage}%
		}
		\subfigure[]{
			\begin{minipage}[t]{0.48\textwidth}
				\centering
				\includegraphics[width=\textwidth]{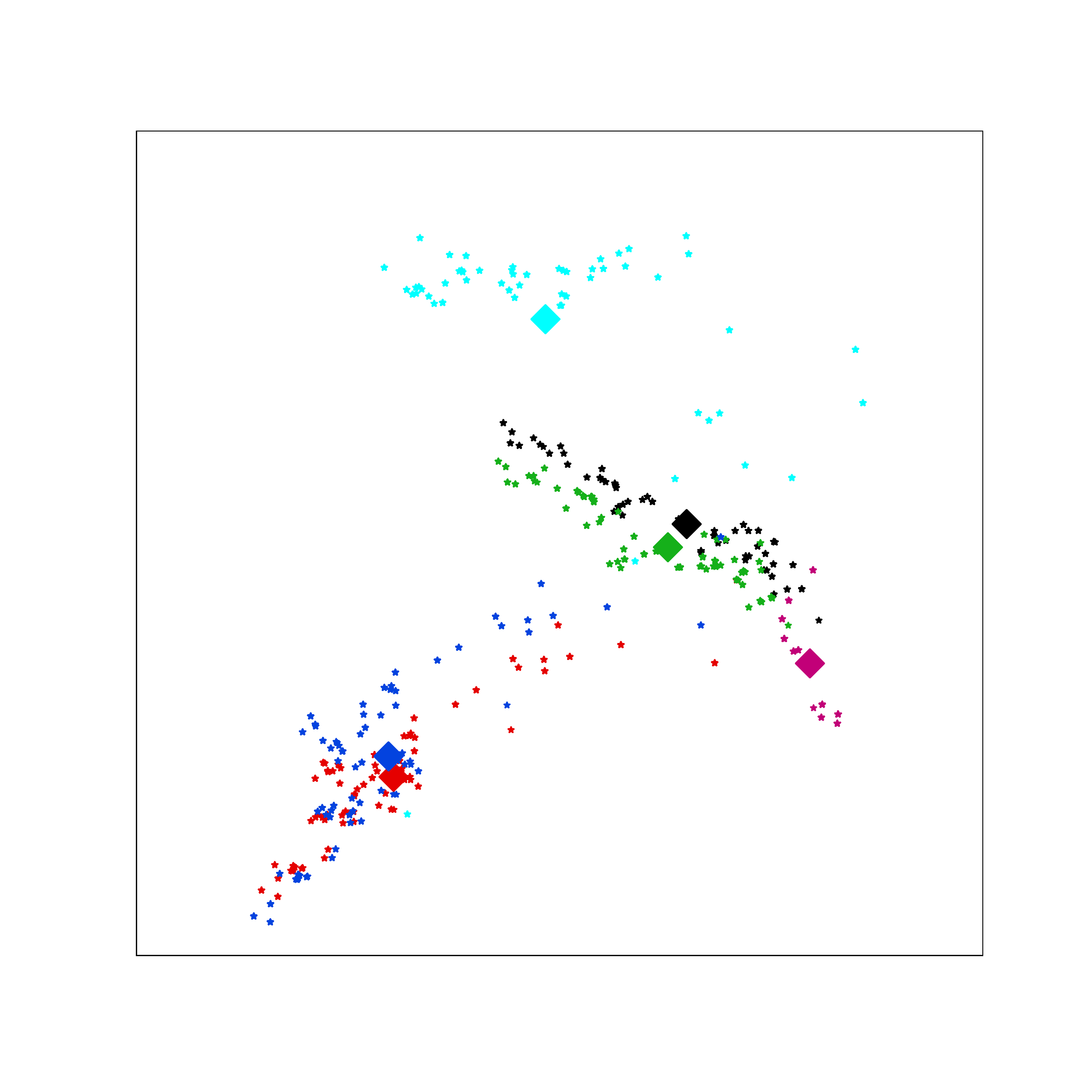}
				
			\end{minipage}%
			
		}%
		\caption{Visualization of purification characteristics. Different color represent the representation results of different types of feedback in FP module, each $\star$ represents the specific feedback generated by the user and item, and 
			${\blacklozenge}$ represents the center of the interest representation cluster.}\label{fig4-5-1}
	\end{minipage}
	\\\\\\
\end{minipage}
of the comparison methods, and we can find that DUMN$_{AF}$ that uses all types of the user's feedback information for interest characterization has better performance than the DUMN$_{IF}$ that only use implicit feedback. Furthermore, DUMN which fine-grained description interest of users with various feedback can further improve the recommendation performance when compared with DUMN$_{AF}$. 
It is intuitive that 
the user preferences indicated 
by different feedback are inconsistent, and the corresponding representation space should also be different. 
Accordingly, DUMN$_{AF}$ without fine-grained modeling of interest will lead to the relative confusion of multi-interest representation.


\subsection{Parameter Sensitivity (RQ3)}\label{section4.4}

In this subsection, we investigate the impact of two hyper-parameters in our developed DUMN framework: the number of slots $m$ and dimension $Z$ of slots. For conciseness, we report experimental results on Alibaba dataset.

We make the number of slots $m$ in DUMN varies in a range of \{16, 32, 64, 128, 256\} and let the dimension $Z$ of each slot varies in a range of \{16, 32, 64, 128, 256\}. From Fig. \ref{fig-4.4}, we can find that when $m$ and $Z$ in each memory network are 256 and 64 respectively, the performance of DUMN is the best. When $m$ and $Z$ are too large or too small, the effect of the model will be worse. The potential explanation for this phenomenon is that the parameters $m$ and $Z$ can directly reflect the storage capacity of the memory network. When the values of these parameters are small, the storage capacity is too weak to effectively store the long-term interests of users. If the values are too large, it may lead to overfitting.

\subsection{Dislike Prediction}
The above experiments show that the user's click preferences can be effectively predicted through DUMN, which is to predict the items that the user is interested in. In this subsection, we further use DUMN to do the prediction task that users are not interested in, which is also a new prediction task recently proposed in DFN. It is dedicated to predicting whether users rate the target item as dislike, so as to avoid the model's prediction results from frustrating users. Table \ref{table4.5} records the experimental results of DUMN and DFN, from which we can find that DUMN can achieve better prediction results than the baseline experimental result, which indicates that our DUMN is more sensitive to the capture of disliked preferences, and the better experimental results come from our fine-grained preference modeling of user feedback.

\begin{minipage}[ht]{\textwidth}
	\begin{minipage}[ht]{0.48\textwidth}
		\makeatletter\def\@captype{figure}\makeatother\subfigure[]{
			\begin{minipage}[t]{0.48\textwidth}
				\centering
				\includegraphics[width=\textwidth]{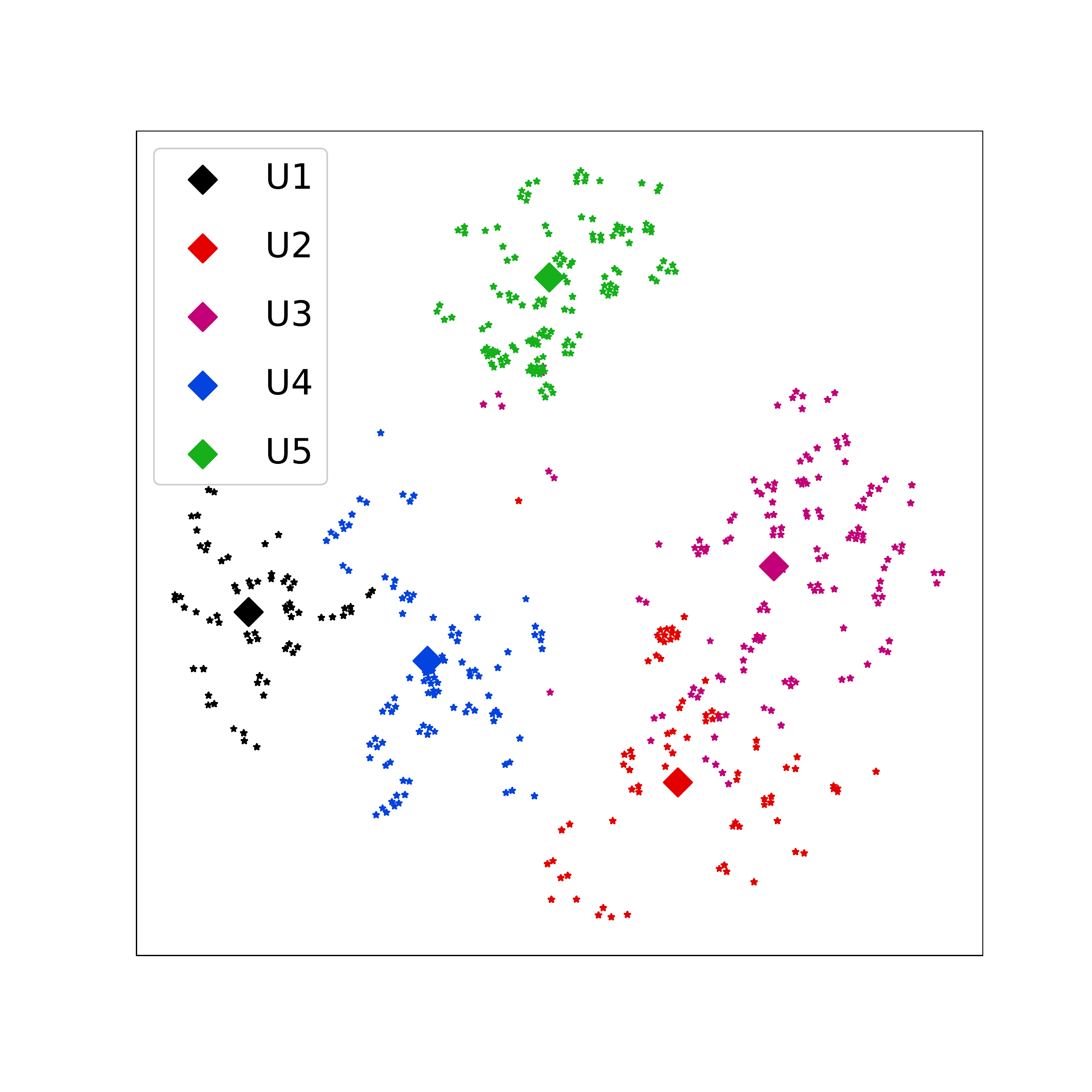}
				
			\end{minipage}%
		}
		\subfigure[]{
			\begin{minipage}[t]{0.48\textwidth}
				\centering
				\includegraphics[width=\textwidth]{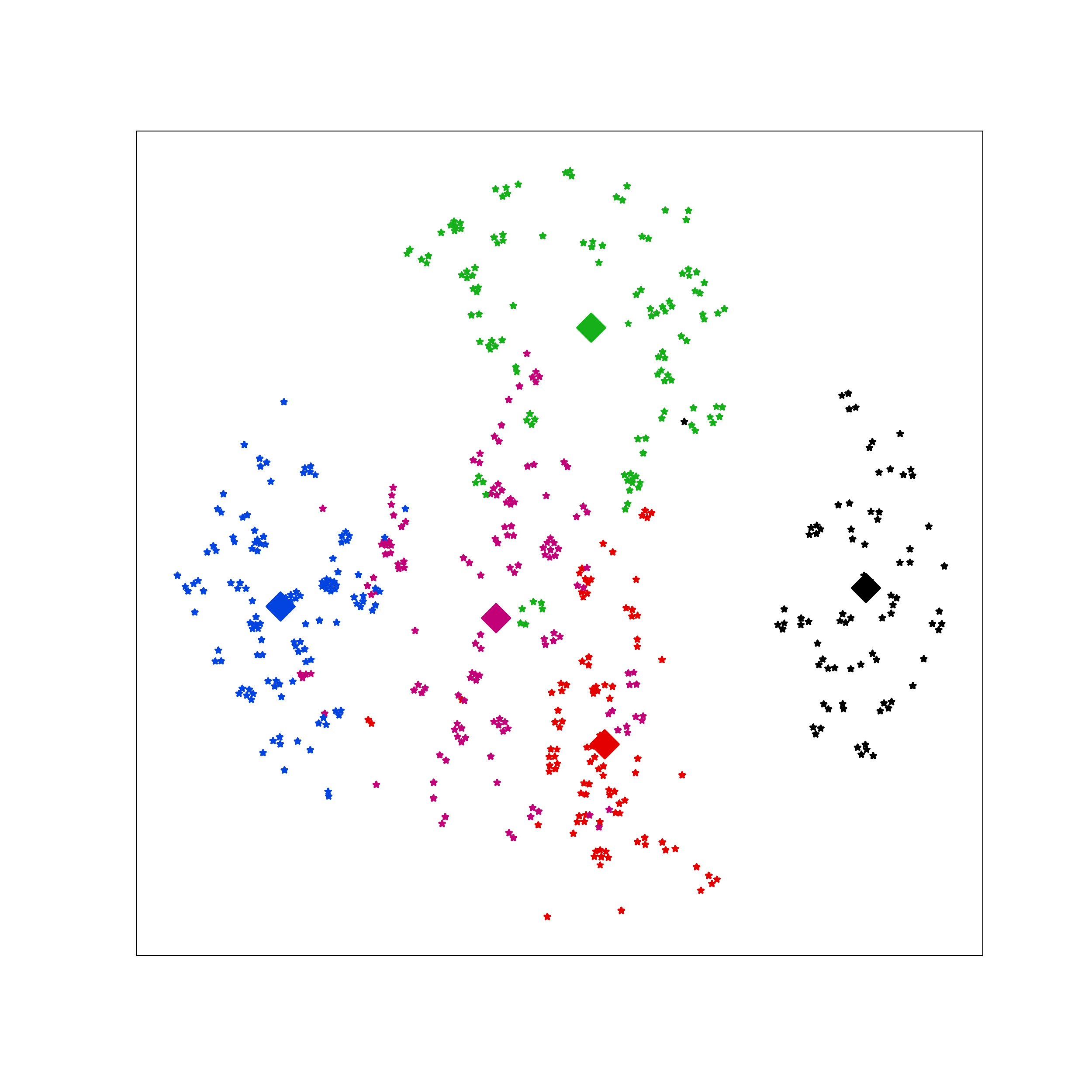}
				
			\end{minipage}%
			
		}%
		\caption{Visualization of long-term interest. Different colors represent the long-term interest representation of different users, and each point refers to the user's long-term interest representation 
			read from the memory network. }\label{fig4-5-2}
	\end{minipage}
	\hspace{.15in}
	\begin{minipage}[ht]{0.48\textwidth}
		\centering
		\makeatletter\def\@captype{figure}\makeatother\subfigure[]{
			\begin{minipage}[t]{0.48\textwidth}
				\centering
				\includegraphics[width=\textwidth]{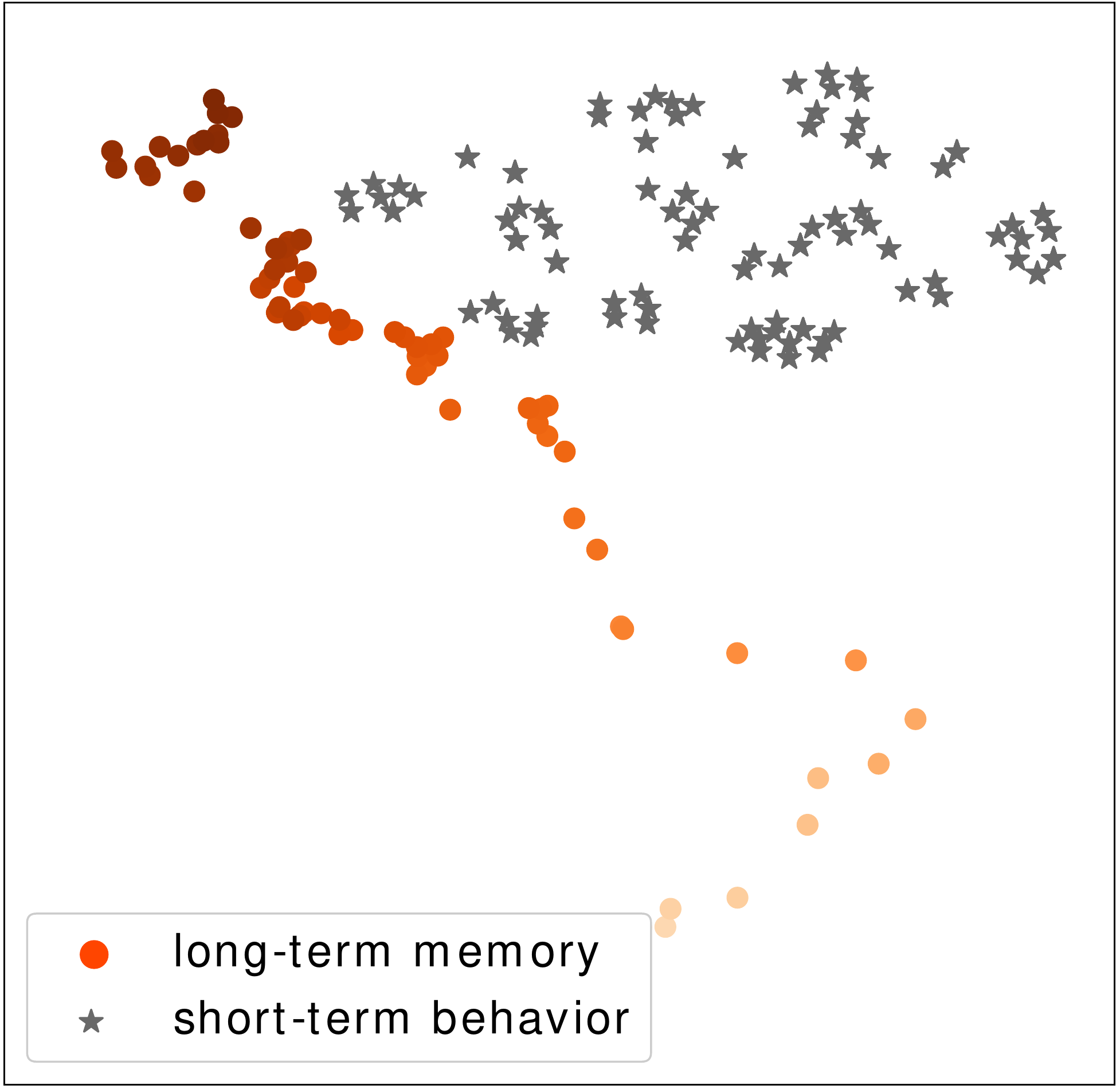}
				
			\end{minipage}%
		}
		\subfigure[]{
			\begin{minipage}[t]{0.48\textwidth}
				\centering
				\includegraphics[width=\textwidth]{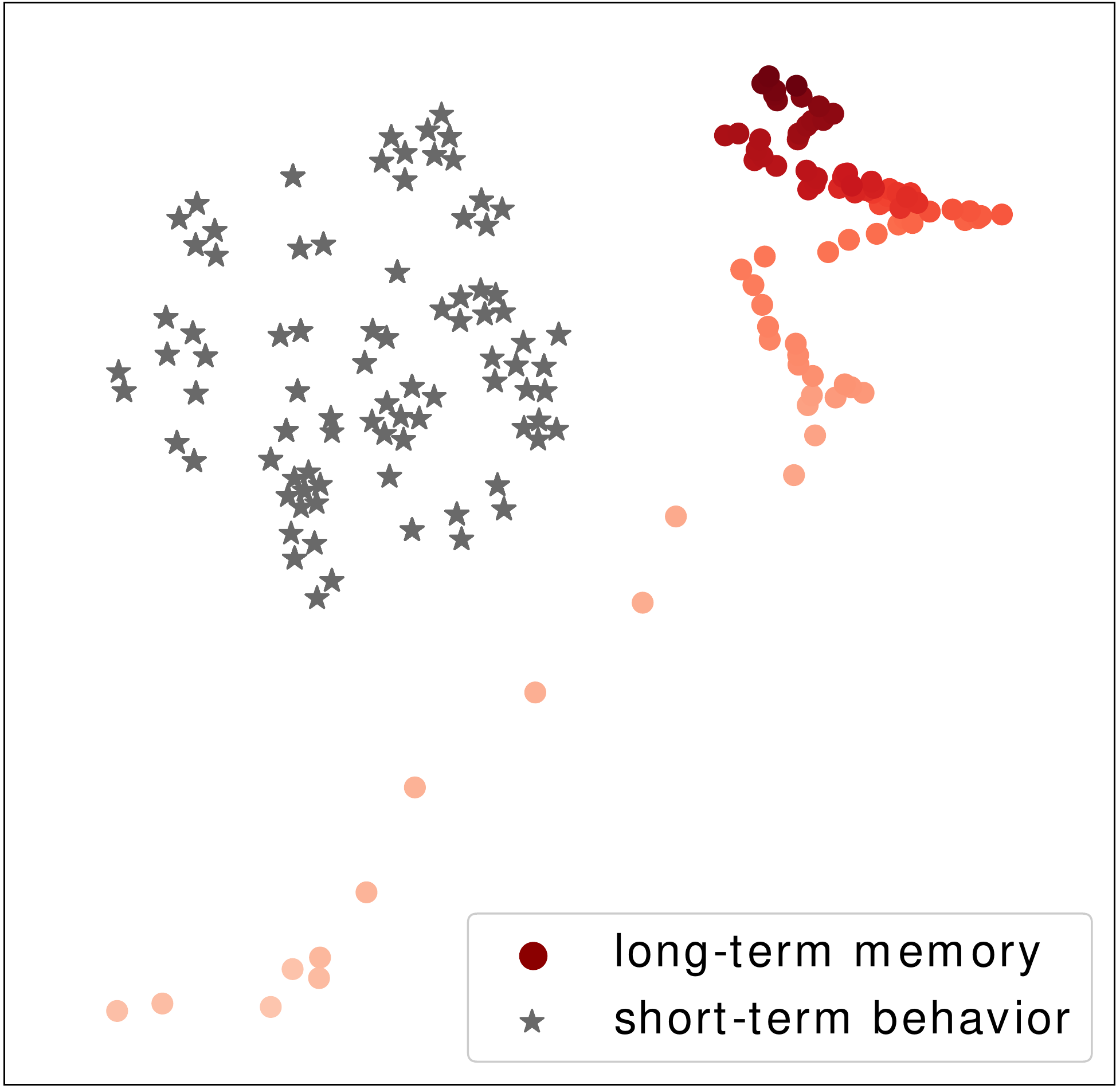}
				
			\end{minipage}%
			
		}%
		\caption{Visualization of long-term and short-term interests. $\bullet$ represents the user's long-term interest, and the deepening of color represents the change process of user's long-term interest in the training process. $\star$ denotes the user's short-term interest. }\label{fig4-5-3}
	\end{minipage}
	\\\\
\end{minipage}

\subsection{Case Study}
In this section, we conduct case study 
to prove that the FP module FP of purifying implicit feedback representation and the PAIR module of capturing long-term interest is effective.

\subsubsection{Display of purification characteristics}First, we show the dimensionality reduction visualization of different short-term interest representations of the same user in Fig. \ref{fig4-5-1}. Specifically, it includes the click and the unclick representations before and after denoising, the like representation and the dislike representation. From the figure, we can find that the noise in implicit feedback representation is effectively removed, and the distance between the denoised representation 
and the representation of orthogonal mapping corresponding to explicit feedback is also far away. For example, compared with before purification, the presentation of click and unclick after purification is far away from that of dislike and like respectively.
%
%
%
%


\subsubsection{Display of long-term interest}Then, we visualized the representation of long-term interests obtained by multiple users from the memory network. It can be observed from Fig. \ref{fig4-5-2} that the long-term interests of the same user are aggregated in the whole representation space, reflecting the stability of long-term interests, and the representations of long-term interests of different users can show intersection or no intersection. Among them, the non-intersection part reflects that users' interests are independent, 
and the intersection part indicates that users' long-term interests are partially similar.
%
%
%
%
\subsubsection{Display of long-term and short-term interests}Last, we visualize the long-term interest representation and short-term interest representation in Fig. \ref{fig4-5-3}, and we can draw the following conclusions: 1) In the training process of DUMN, the user's long-term interest representation is gradually aggregated into a smaller space, reaching a more stable state; and 2) Compared with the relatively stable long-term interest representation learned, the short-term interest representation of the user presents a more dispersed state under the unified feature space, which reflects the difference between long-term and short-term interests. It is worth noting that our DUMN model can effectively capture users' long-term and short-term interests of users through the FP module and the UMN module.
%
%
%
%
\section{Conclusion}In this paper, we propose a novel denoising user-aware memory network (DUMN), which constructs four feedback sequences of users, namely click, unclick, like and dislike, to model users' preferences in a fine-grained way. DUMN uses feature purification layer and user memory network layer to purify the implicit feedback sequence representation of users and capture the stable long-term preference of users, respectively. A large number of experiments verify the effectiveness of the proposed model in the CTR prediction task. Experiments verify the effectiveness of the proposed model in the CTR prediction task. In the future, we intend to further explore the impact of noise purification and long-term interest in different recommendation scenarios, and subdivide the fine-grained representation of users on the basis of the existing to obtain more comprehensive unbiased preference representation of users.

\bibliographystyle{ACM-Reference-Format}
\bibliography{sample-sigconf}


\begin{thebibliography}{48}


\ifx \showCODEN    \undefined \def \showCODEN     #1{\unskip}     \fi
\ifx \showDOI      \undefined \def \showDOI       #1{#1}\fi
\ifx \showISBNx    \undefined \def \showISBNx     #1{\unskip}     \fi
\ifx \showISBNxiii \undefined \def \showISBNxiii  #1{\unskip}     \fi
\ifx \showISSN     \undefined \def \showISSN      #1{\unskip}     \fi
\ifx \showLCCN     \undefined \def \showLCCN      #1{\unskip}     \fi
\ifx \shownote     \undefined \def \shownote      #1{#1}          \fi
\ifx \showarticletitle \undefined \def \showarticletitle #1{#1}   \fi
\ifx \showURL      \undefined \def \showURL       {\relax}        \fi
\providecommand\bibfield[2]{#2}
\providecommand\bibinfo[2]{#2}
\providecommand\natexlab[1]{#1}
\providecommand\showeprint[2][]{arXiv:#2}

\bibitem[\protect\citeauthoryear{Borisyuk, Zhang, and Kenthapadi}{Borisyuk
  et~al\mbox{.}}{2017}]%
        {borisyuk2017lijar}
\bibfield{author}{\bibinfo{person}{Fedor Borisyuk}, \bibinfo{person}{Liang
  Zhang}, {and} \bibinfo{person}{Krishnaram Kenthapadi}.}
  \bibinfo{year}{2017}\natexlab{}.
\newblock \showarticletitle{LiJAR: A system for job application redistribution
  towards efficient career marketplace}. In
  \bibinfo{booktitle}{\emph{Proceedings of the 23rd ACM SIGKDD International
  Conference on Knowledge Discovery and Data Mining}}.
  \bibinfo{pages}{1397--1406}.
\newblock


\bibitem[\protect\citeauthoryear{Chen, Wang, Zhou, Shi, Chen, Feng, and
  Chen}{Chen et~al\mbox{.}}{2020}]%
        {chen2020fast}
\bibfield{author}{\bibinfo{person}{Jiawei Chen}, \bibinfo{person}{Can Wang},
  \bibinfo{person}{Sheng Zhou}, \bibinfo{person}{Qihao Shi},
  \bibinfo{person}{Jingbang Chen}, \bibinfo{person}{Yan Feng}, {and}
  \bibinfo{person}{Chun Chen}.} \bibinfo{year}{2020}\natexlab{}.
\newblock \showarticletitle{Fast adaptively weighted matrix factorization for
  recommendation with implicit feedback}. In
  \bibinfo{booktitle}{\emph{Proceedings of the AAAI Conference on Artificial
  Intelligence}}, Vol.~\bibinfo{volume}{34}. \bibinfo{pages}{3470--3477}.
\newblock


\bibitem[\protect\citeauthoryear{Cheng, Koc, Harmsen, Shaked, Chandra, Aradhye,
  Anderson, Corrado, Chai, Ispir, et~al\mbox{.}}{Cheng et~al\mbox{.}}{2016}]%
        {cheng2016wide}
\bibfield{author}{\bibinfo{person}{Heng-Tze Cheng}, \bibinfo{person}{Levent
  Koc}, \bibinfo{person}{Jeremiah Harmsen}, \bibinfo{person}{Tal Shaked},
  \bibinfo{person}{Tushar Chandra}, \bibinfo{person}{Hrishi Aradhye},
  \bibinfo{person}{Glen Anderson}, \bibinfo{person}{Greg Corrado},
  \bibinfo{person}{Wei Chai}, \bibinfo{person}{Mustafa Ispir}, {et~al\mbox{.}}}
  \bibinfo{year}{2016}\natexlab{}.
\newblock \showarticletitle{Wide \& deep learning for recommender systems}. In
  \bibinfo{booktitle}{\emph{Proceedings of the 1st workshop on deep learning
  for recommender systems}}. \bibinfo{pages}{7--10}.
\newblock


\bibitem[\protect\citeauthoryear{Covington, Adams, and Sargin}{Covington
  et~al\mbox{.}}{2016}]%
        {covington2016deep}
\bibfield{author}{\bibinfo{person}{Paul Covington}, \bibinfo{person}{Jay
  Adams}, {and} \bibinfo{person}{Emre Sargin}.}
  \bibinfo{year}{2016}\natexlab{}.
\newblock \showarticletitle{Deep neural networks for youtube recommendations}.
  In \bibinfo{booktitle}{\emph{Proceedings of the 10th ACM conference on
  recommender systems}}. \bibinfo{pages}{191--198}.
\newblock


\bibitem[\protect\citeauthoryear{Devlin, Chang, Lee, and Toutanova}{Devlin
  et~al\mbox{.}}{2019}]%
        {devlin2019bert}
\bibfield{author}{\bibinfo{person}{Jacob Devlin}, \bibinfo{person}{Ming-Wei
  Chang}, \bibinfo{person}{Kenton Lee}, {and} \bibinfo{person}{Kristina
  Toutanova}.} \bibinfo{year}{2019}\natexlab{}.
\newblock \showarticletitle{BERT: Pre-training of Deep Bidirectional
  Transformers for Language Understanding}. In
  \bibinfo{booktitle}{\emph{NAACL-HLT (1)}}.
\newblock


\bibitem[\protect\citeauthoryear{Feng, Lv, Shen, Wang, Sun, Zhu, and Yang}{Feng
  et~al\mbox{.}}{2019}]%
        {DBLP:conf/ijcai/FengLSWSZY19}
\bibfield{author}{\bibinfo{person}{Yufei Feng}, \bibinfo{person}{Fuyu Lv},
  \bibinfo{person}{Weichen Shen}, \bibinfo{person}{Menghan Wang},
  \bibinfo{person}{Fei Sun}, \bibinfo{person}{Yu Zhu}, {and}
  \bibinfo{person}{Keping Yang}.} \bibinfo{year}{2019}\natexlab{}.
\newblock \showarticletitle{Deep Session Interest Network for Click-Through
  Rate Prediction}. In \bibinfo{booktitle}{\emph{IJCAI International Joint
  Conference on Artificial Intelligence}},
  \bibfield{editor}{\bibinfo{person}{Sarit Kraus}} (Ed.).
  \bibinfo{publisher}{ijcai.org}, \bibinfo{pages}{2301--2307}.
\newblock


\bibitem[\protect\citeauthoryear{F{\'e}votte and Idier}{F{\'e}votte and
  Idier}{2011}]%
        {fevotte2011algorithms}
\bibfield{author}{\bibinfo{person}{C{\'e}dric F{\'e}votte} {and}
  \bibinfo{person}{J{\'e}r{\^o}me Idier}.} \bibinfo{year}{2011}\natexlab{}.
\newblock \showarticletitle{Algorithms for nonnegative matrix factorization
  with the $\beta$-divergence}.
\newblock \bibinfo{journal}{\emph{Neural computation}} \bibinfo{volume}{23},
  \bibinfo{number}{9} (\bibinfo{year}{2011}), \bibinfo{pages}{2421--2456}.
\newblock


\bibitem[\protect\citeauthoryear{Goldberg, Nichols, Oki, and Terry}{Goldberg
  et~al\mbox{.}}{1992}]%
        {goldberg1992using}
\bibfield{author}{\bibinfo{person}{David Goldberg}, \bibinfo{person}{David
  Nichols}, \bibinfo{person}{Brian~M Oki}, {and} \bibinfo{person}{Douglas
  Terry}.} \bibinfo{year}{1992}\natexlab{}.
\newblock \showarticletitle{Using collaborative filtering to weave an
  information tapestry}.
\newblock \bibinfo{journal}{\emph{Commun. ACM}} \bibinfo{volume}{35},
  \bibinfo{number}{12} (\bibinfo{year}{1992}), \bibinfo{pages}{61--70}.
\newblock


\bibitem[\protect\citeauthoryear{Graves, Wayne, and Danihelka}{Graves
  et~al\mbox{.}}{2014}]%
        {graves2014neural}
\bibfield{author}{\bibinfo{person}{Alex Graves}, \bibinfo{person}{Greg Wayne},
  {and} \bibinfo{person}{Ivo Danihelka}.} \bibinfo{year}{2014}\natexlab{}.
\newblock \showarticletitle{Neural turing machines}.
\newblock \bibinfo{journal}{\emph{arXiv preprint arXiv:1410.5401}}
  (\bibinfo{year}{2014}).
\newblock


\bibitem[\protect\citeauthoryear{Gu, Ding, Wang, Zou, Liu, and Yin}{Gu
  et~al\mbox{.}}{2020}]%
        {gu2020deep}
\bibfield{author}{\bibinfo{person}{Yulong Gu}, \bibinfo{person}{Zhuoye Ding},
  \bibinfo{person}{Shuaiqiang Wang}, \bibinfo{person}{Lixin Zou},
  \bibinfo{person}{Yiding Liu}, {and} \bibinfo{person}{Dawei Yin}.}
  \bibinfo{year}{2020}\natexlab{}.
\newblock \showarticletitle{Deep Multifaceted Transformers for Multi-objective
  Ranking in Large-Scale E-commerce Recommender Systems}. In
  \bibinfo{booktitle}{\emph{Proceedings of the 29th ACM International
  Conference on Information \& Knowledge Management}}.
  \bibinfo{pages}{2493--2500}.
\newblock


\bibitem[\protect\citeauthoryear{Guo, Tang, Ye, Li, and He}{Guo
  et~al\mbox{.}}{2017}]%
        {guo2017deepfm}
\bibfield{author}{\bibinfo{person}{Huifeng Guo}, \bibinfo{person}{Ruiming
  Tang}, \bibinfo{person}{Yunming Ye}, \bibinfo{person}{Zhenguo Li}, {and}
  \bibinfo{person}{Xiuqiang He}.} \bibinfo{year}{2017}\natexlab{}.
\newblock \showarticletitle{DeepFM: a factorization-machine based neural
  network for CTR prediction}.
\newblock \bibinfo{journal}{\emph{arXiv preprint arXiv:1703.04247}}
  (\bibinfo{year}{2017}).
\newblock


\bibitem[\protect\citeauthoryear{Hadash, Shalom, and Osadchy}{Hadash
  et~al\mbox{.}}{2018}]%
        {hadash2018rank}
\bibfield{author}{\bibinfo{person}{Guy Hadash}, \bibinfo{person}{Oren~Sar
  Shalom}, {and} \bibinfo{person}{Rita Osadchy}.}
  \bibinfo{year}{2018}\natexlab{}.
\newblock \showarticletitle{Rank and rate: multi-task learning for recommender
  systems}. In \bibinfo{booktitle}{\emph{Proceedings of the 12th ACM Conference
  on Recommender Systems}}. \bibinfo{pages}{451--454}.
\newblock


\bibitem[\protect\citeauthoryear{Hidasi, Karatzoglou, Baltrunas, and
  Tikk}{Hidasi et~al\mbox{.}}{2016}]%
        {DBLP:journals/corr/HidasiKBT15}
\bibfield{author}{\bibinfo{person}{Bal{\'{a}}zs Hidasi},
  \bibinfo{person}{Alexandros Karatzoglou}, \bibinfo{person}{Linas Baltrunas},
  {and} \bibinfo{person}{Domonkos Tikk}.} \bibinfo{year}{2016}\natexlab{}.
\newblock \showarticletitle{Session-based Recommendations with Recurrent Neural
  Networks}. In \bibinfo{booktitle}{\emph{4th International Conference on
  Learning Representations (ICLR)}}.
\newblock


\bibitem[\protect\citeauthoryear{Hu, Koren, and Volinsky}{Hu
  et~al\mbox{.}}{2008}]%
        {hu2008collaborative}
\bibfield{author}{\bibinfo{person}{Yifan Hu}, \bibinfo{person}{Yehuda Koren},
  {and} \bibinfo{person}{Chris Volinsky}.} \bibinfo{year}{2008}\natexlab{}.
\newblock \showarticletitle{Collaborative filtering for implicit feedback
  datasets}. In \bibinfo{booktitle}{\emph{2008 Eighth IEEE International
  Conference on Data Mining}}. Ieee, \bibinfo{pages}{263--272}.
\newblock


\bibitem[\protect\citeauthoryear{Huang, Liu, Van Der~Maaten, and
  Weinberger}{Huang et~al\mbox{.}}{2017}]%
        {huang2017densely}
\bibfield{author}{\bibinfo{person}{Gao Huang}, \bibinfo{person}{Zhuang Liu},
  \bibinfo{person}{Laurens Van Der~Maaten}, {and} \bibinfo{person}{Kilian~Q
  Weinberger}.} \bibinfo{year}{2017}\natexlab{}.
\newblock \showarticletitle{Densely connected convolutional networks}. In
  \bibinfo{booktitle}{\emph{Proceedings of the IEEE conference on computer
  vision and pattern recognition}}. \bibinfo{pages}{4700--4708}.
\newblock


\bibitem[\protect\citeauthoryear{Jadidinejad, Macdonald, and Ounis}{Jadidinejad
  et~al\mbox{.}}{2019}]%
        {jadidinejad2019unifying}
\bibfield{author}{\bibinfo{person}{Amir~H Jadidinejad}, \bibinfo{person}{Craig
  Macdonald}, {and} \bibinfo{person}{Iadh Ounis}.}
  \bibinfo{year}{2019}\natexlab{}.
\newblock \showarticletitle{Unifying explicit and implicit feedback for rating
  prediction and ranking recommendation tasks}. In
  \bibinfo{booktitle}{\emph{Proceedings of the 2019 ACM SIGIR International
  Conference on Theory of Information Retrieval}}. \bibinfo{pages}{149--156}.
\newblock


\bibitem[\protect\citeauthoryear{Jin, Qin, Fang, Du, Zhang, Yu, Zhang, and
  Smola}{Jin et~al\mbox{.}}{2020}]%
        {jin2020efficient}
\bibfield{author}{\bibinfo{person}{Jiarui Jin}, \bibinfo{person}{Jiarui Qin},
  \bibinfo{person}{Yuchen Fang}, \bibinfo{person}{Kounianhua Du},
  \bibinfo{person}{Weinan Zhang}, \bibinfo{person}{Yong Yu},
  \bibinfo{person}{Zheng Zhang}, {and} \bibinfo{person}{Alexander~J Smola}.}
  \bibinfo{year}{2020}\natexlab{}.
\newblock \showarticletitle{An Efficient Neighborhood-based Interaction Model
  for Recommendation on Heterogeneous Graph}. In
  \bibinfo{booktitle}{\emph{Proceedings of the 26th ACM SIGKDD International
  Conference on Knowledge Discovery \& Data Mining}}. \bibinfo{pages}{75--84}.
\newblock


\bibitem[\protect\citeauthoryear{Kang and McAuley}{Kang and McAuley}{2018}]%
        {kang2018self}
\bibfield{author}{\bibinfo{person}{Wang-Cheng Kang} {and}
  \bibinfo{person}{Julian McAuley}.} \bibinfo{year}{2018}\natexlab{}.
\newblock \showarticletitle{Self-attentive sequential recommendation}. In
  \bibinfo{booktitle}{\emph{2018 IEEE International Conference on Data Mining
  (ICDM)}}. IEEE, \bibinfo{pages}{197--206}.
\newblock


\bibitem[\protect\citeauthoryear{Kipf and Welling}{Kipf and Welling}{2016}]%
        {kipf2016semi}
\bibfield{author}{\bibinfo{person}{Thomas~N Kipf} {and} \bibinfo{person}{Max
  Welling}.} \bibinfo{year}{2016}\natexlab{}.
\newblock \showarticletitle{Semi-supervised classification with graph
  convolutional networks}.
\newblock \bibinfo{journal}{\emph{arXiv preprint arXiv:1609.02907}}
  (\bibinfo{year}{2016}).
\newblock


\bibitem[\protect\citeauthoryear{Koren}{Koren}{2008}]%
        {koren2008factorization}
\bibfield{author}{\bibinfo{person}{Yehuda Koren}.}
  \bibinfo{year}{2008}\natexlab{}.
\newblock \showarticletitle{Factorization meets the neighborhood: a
  multifaceted collaborative filtering model}. In
  \bibinfo{booktitle}{\emph{Proceedings of the 14th ACM SIGKDD international
  conference on Knowledge discovery and data mining}}.
  \bibinfo{pages}{426--434}.
\newblock


\bibitem[\protect\citeauthoryear{Koren, Bell, and Volinsky}{Koren
  et~al\mbox{.}}{2009}]%
        {koren2009matrix}
\bibfield{author}{\bibinfo{person}{Yehuda Koren}, \bibinfo{person}{Robert
  Bell}, {and} \bibinfo{person}{Chris Volinsky}.}
  \bibinfo{year}{2009}\natexlab{}.
\newblock \showarticletitle{Matrix factorization techniques for recommender
  systems}.
\newblock \bibinfo{journal}{\emph{Computer}} \bibinfo{volume}{42},
  \bibinfo{number}{8} (\bibinfo{year}{2009}), \bibinfo{pages}{30--37}.
\newblock


\bibitem[\protect\citeauthoryear{Li, Ren, Chen, Ren, Lian, and Ma}{Li
  et~al\mbox{.}}{2017}]%
        {li2017neural}
\bibfield{author}{\bibinfo{person}{Jing Li}, \bibinfo{person}{Pengjie Ren},
  \bibinfo{person}{Zhumin Chen}, \bibinfo{person}{Zhaochun Ren},
  \bibinfo{person}{Tao Lian}, {and} \bibinfo{person}{Jun Ma}.}
  \bibinfo{year}{2017}\natexlab{}.
\newblock \showarticletitle{Neural attentive session-based recommendation}. In
  \bibinfo{booktitle}{\emph{Proceedings of the 2017 ACM on Conference on
  Information and Knowledge Management}}. \bibinfo{pages}{1419--1428}.
\newblock


\bibitem[\protect\citeauthoryear{Li, Cheng, Chen, Chen, and Wang}{Li
  et~al\mbox{.}}{2020}]%
        {li2020interpretable}
\bibfield{author}{\bibinfo{person}{Zeyu Li}, \bibinfo{person}{Wei Cheng},
  \bibinfo{person}{Yang Chen}, \bibinfo{person}{Haifeng Chen}, {and}
  \bibinfo{person}{Wei Wang}.} \bibinfo{year}{2020}\natexlab{}.
\newblock \showarticletitle{Interpretable Click-Through Rate Prediction through
  Hierarchical Attention}. In \bibinfo{booktitle}{\emph{Proceedings of the 13th
  International Conference on Web Search and Data Mining}}.
  \bibinfo{pages}{313--321}.
\newblock


\bibitem[\protect\citeauthoryear{Lian, Zhou, Zhang, Chen, Xie, and Sun}{Lian
  et~al\mbox{.}}{2018}]%
        {lian2018xdeepfm}
\bibfield{author}{\bibinfo{person}{Jianxun Lian}, \bibinfo{person}{Xiaohuan
  Zhou}, \bibinfo{person}{Fuzheng Zhang}, \bibinfo{person}{Zhongxia Chen},
  \bibinfo{person}{Xing Xie}, {and} \bibinfo{person}{Guangzhong Sun}.}
  \bibinfo{year}{2018}\natexlab{}.
\newblock \showarticletitle{xdeepfm: Combining explicit and implicit feature
  interactions for recommender systems}. In
  \bibinfo{booktitle}{\emph{Proceedings of the 24th ACM SIGKDD International
  Conference on Knowledge Discovery \& Data Mining}}.
  \bibinfo{pages}{1754--1763}.
\newblock


\bibitem[\protect\citeauthoryear{Liu, Xiang, Zhao, and Yang}{Liu
  et~al\mbox{.}}{2010}]%
        {liu2010unifying}
\bibfield{author}{\bibinfo{person}{Nathan~N Liu}, \bibinfo{person}{Evan~W
  Xiang}, \bibinfo{person}{Min Zhao}, {and} \bibinfo{person}{Qiang Yang}.}
  \bibinfo{year}{2010}\natexlab{}.
\newblock \showarticletitle{Unifying explicit and implicit feedback for
  collaborative filtering}. In \bibinfo{booktitle}{\emph{Proceedings of the
  19th ACM international conference on Information and knowledge management}}.
  \bibinfo{pages}{1445--1448}.
\newblock


\bibitem[\protect\citeauthoryear{Lv, Jin, Yu, Sun, Lin, Yang, and Ng}{Lv
  et~al\mbox{.}}{2019}]%
        {lv2019sdm}
\bibfield{author}{\bibinfo{person}{Fuyu Lv}, \bibinfo{person}{Taiwei Jin},
  \bibinfo{person}{Changlong Yu}, \bibinfo{person}{Fei Sun},
  \bibinfo{person}{Quan Lin}, \bibinfo{person}{Keping Yang}, {and}
  \bibinfo{person}{Wilfred Ng}.} \bibinfo{year}{2019}\natexlab{}.
\newblock \showarticletitle{SDM: Sequential deep matching model for online
  large-scale recommender system}. In \bibinfo{booktitle}{\emph{Proceedings of
  the 28th ACM International Conference on Information and Knowledge
  Management}}. \bibinfo{pages}{2635--2643}.
\newblock


\bibitem[\protect\citeauthoryear{Lv, Li, Guo, Yu, Sun, Jin, and Yang}{Lv
  et~al\mbox{.}}{2020}]%
        {lv2020unclicked}
\bibfield{author}{\bibinfo{person}{Fuyu Lv}, \bibinfo{person}{Mengxue Li},
  \bibinfo{person}{Tonglei Guo}, \bibinfo{person}{Changlong Yu},
  \bibinfo{person}{Fei Sun}, \bibinfo{person}{Taiwei Jin}, {and}
  \bibinfo{person}{Keping Yang}.} \bibinfo{year}{2020}\natexlab{}.
\newblock \showarticletitle{Unclicked User Behaviors Enhanced
  SequentialRecommendation}.
\newblock \bibinfo{journal}{\emph{arXiv preprint arXiv:2010.12837}}
  (\bibinfo{year}{2020}).
\newblock


\bibitem[\protect\citeauthoryear{Ni, Ou, Liu, Li, Ou, Zeng, and Si}{Ni
  et~al\mbox{.}}{2018}]%
        {ni2018perceive}
\bibfield{author}{\bibinfo{person}{Yabo Ni}, \bibinfo{person}{Dan Ou},
  \bibinfo{person}{Shichen Liu}, \bibinfo{person}{Xiang Li},
  \bibinfo{person}{Wenwu Ou}, \bibinfo{person}{Anxiang Zeng}, {and}
  \bibinfo{person}{Luo Si}.} \bibinfo{year}{2018}\natexlab{}.
\newblock \showarticletitle{Perceive your users in depth: Learning universal
  user representations from multiple e-commerce tasks}. In
  \bibinfo{booktitle}{\emph{Proceedings of the 24th ACM SIGKDD International
  Conference on Knowledge Discovery \& Data Mining}}.
  \bibinfo{pages}{596--605}.
\newblock


\bibitem[\protect\citeauthoryear{Oh, Lee, Lim, and Choi}{Oh
  et~al\mbox{.}}{2014}]%
        {2014Personalized}
\bibfield{author}{\bibinfo{person}{Kyo~Joong Oh}, \bibinfo{person}{Won~Jo Lee},
  \bibinfo{person}{Chae~Gyun Lim}, {and} \bibinfo{person}{Ho~Jin Choi}.}
  \bibinfo{year}{2014}\natexlab{}.
\newblock \bibinfo{booktitle}{\emph{Personalized news recommendation using
  classified keywords to capture user preference}}.
\newblock


\bibitem[\protect\citeauthoryear{Pi, Bian, Zhou, Zhu, and Gai}{Pi
  et~al\mbox{.}}{2019}]%
        {pi2019practice}
\bibfield{author}{\bibinfo{person}{Qi Pi}, \bibinfo{person}{Weijie Bian},
  \bibinfo{person}{Guorui Zhou}, \bibinfo{person}{Xiaoqiang Zhu}, {and}
  \bibinfo{person}{Kun Gai}.} \bibinfo{year}{2019}\natexlab{}.
\newblock \showarticletitle{Practice on long sequential user behavior modeling
  for click-through rate prediction}. In \bibinfo{booktitle}{\emph{Proceedings
  of the 25th ACM SIGKDD International Conference on Knowledge Discovery \&
  Data Mining}}. \bibinfo{pages}{2671--2679}.
\newblock


\bibitem[\protect\citeauthoryear{Qin, Hu, and Liu}{Qin et~al\mbox{.}}{2020}]%
        {qin2020feature}
\bibfield{author}{\bibinfo{person}{Qi Qin}, \bibinfo{person}{Wenpeng Hu}, {and}
  \bibinfo{person}{Bing Liu}.} \bibinfo{year}{2020}\natexlab{}.
\newblock \showarticletitle{Feature projection for improved text
  classification}. In \bibinfo{booktitle}{\emph{Proceedings of the 58th Annual
  Meeting of the Association for Computational Linguistics}}.
  \bibinfo{pages}{8161--8171}.
\newblock


\bibitem[\protect\citeauthoryear{Qu, Cai, Ren, Zhang, Yu, Wen, and Wang}{Qu
  et~al\mbox{.}}{2016}]%
        {qu2016product}
\bibfield{author}{\bibinfo{person}{Yanru Qu}, \bibinfo{person}{Han Cai},
  \bibinfo{person}{Kan Ren}, \bibinfo{person}{Weinan Zhang},
  \bibinfo{person}{Yong Yu}, \bibinfo{person}{Ying Wen}, {and}
  \bibinfo{person}{Jun Wang}.} \bibinfo{year}{2016}\natexlab{}.
\newblock \showarticletitle{Product-based neural networks for user response
  prediction}. In \bibinfo{booktitle}{\emph{2016 IEEE 16th International
  Conference on Data Mining (ICDM)}}. IEEE, \bibinfo{pages}{1149--1154}.
\newblock


\bibitem[\protect\citeauthoryear{Redmon, Divvala, Girshick, and Farhadi}{Redmon
  et~al\mbox{.}}{2016}]%
        {redmon2016you}
\bibfield{author}{\bibinfo{person}{Joseph Redmon}, \bibinfo{person}{Santosh
  Divvala}, \bibinfo{person}{Ross Girshick}, {and} \bibinfo{person}{Ali
  Farhadi}.} \bibinfo{year}{2016}\natexlab{}.
\newblock \showarticletitle{You only look once: Unified, real-time object
  detection}. In \bibinfo{booktitle}{\emph{Proceedings of the IEEE conference
  on computer vision and pattern recognition}}. \bibinfo{pages}{779--788}.
\newblock


\bibitem[\protect\citeauthoryear{Shenbin, Alekseev, Tutubalina, Malykh, and
  Nikolenko}{Shenbin et~al\mbox{.}}{2020}]%
        {shenbin2020recvae}
\bibfield{author}{\bibinfo{person}{Ilya Shenbin}, \bibinfo{person}{Anton
  Alekseev}, \bibinfo{person}{Elena Tutubalina}, \bibinfo{person}{Valentin
  Malykh}, {and} \bibinfo{person}{Sergey~I Nikolenko}.}
  \bibinfo{year}{2020}\natexlab{}.
\newblock \showarticletitle{RecVAE: A new variational autoencoder for Top-N
  recommendations with implicit feedback}. In
  \bibinfo{booktitle}{\emph{Proceedings of the 13th International Conference on
  Web Search and Data Mining}}. \bibinfo{pages}{528--536}.
\newblock


\bibitem[\protect\citeauthoryear{Shi, Hu, Zhao, and Philip}{Shi
  et~al\mbox{.}}{2018}]%
        {shi2018heterogeneous}
\bibfield{author}{\bibinfo{person}{Chuan Shi}, \bibinfo{person}{Binbin Hu},
  \bibinfo{person}{Wayne~Xin Zhao}, {and} \bibinfo{person}{S~Yu Philip}.}
  \bibinfo{year}{2018}\natexlab{}.
\newblock \showarticletitle{Heterogeneous information network embedding for
  recommendation}.
\newblock \bibinfo{journal}{\emph{IEEE Transactions on Knowledge and Data
  Engineering}} \bibinfo{volume}{31}, \bibinfo{number}{2}
  (\bibinfo{year}{2018}), \bibinfo{pages}{357--370}.
\newblock


\bibitem[\protect\citeauthoryear{Song, Shi, Xiao, Duan, Xu, Zhang, and
  Tang}{Song et~al\mbox{.}}{2019}]%
        {song2019autoint}
\bibfield{author}{\bibinfo{person}{Weiping Song}, \bibinfo{person}{Chence Shi},
  \bibinfo{person}{Zhiping Xiao}, \bibinfo{person}{Zhijian Duan},
  \bibinfo{person}{Yewen Xu}, \bibinfo{person}{Ming Zhang}, {and}
  \bibinfo{person}{Jian Tang}.} \bibinfo{year}{2019}\natexlab{}.
\newblock \showarticletitle{Autoint: Automatic feature interaction learning via
  self-attentive neural networks}. In \bibinfo{booktitle}{\emph{Proceedings of
  the 28th ACM International Conference on Information and Knowledge
  Management}}. \bibinfo{pages}{1161--1170}.
\newblock


\bibitem[\protect\citeauthoryear{Tang and Wang}{Tang and Wang}{2018}]%
        {tang2018personalized}
\bibfield{author}{\bibinfo{person}{Jiaxi Tang} {and} \bibinfo{person}{Ke
  Wang}.} \bibinfo{year}{2018}\natexlab{}.
\newblock \showarticletitle{Personalized top-n sequential recommendation via
  convolutional sequence embedding}. In \bibinfo{booktitle}{\emph{Proceedings
  of the Eleventh ACM International Conference on Web Search and Data Mining}}.
  \bibinfo{pages}{565--573}.
\newblock


\bibitem[\protect\citeauthoryear{Van Den~Oord, Dieleman, and Schrauwen}{Van
  Den~Oord et~al\mbox{.}}{2013}]%
        {van2013deep}
\bibfield{author}{\bibinfo{person}{A{\"a}ron Van Den~Oord},
  \bibinfo{person}{Sander Dieleman}, {and} \bibinfo{person}{Benjamin
  Schrauwen}.} \bibinfo{year}{2013}\natexlab{}.
\newblock \showarticletitle{Deep content-based music recommendation}. In
  \bibinfo{booktitle}{\emph{Neural Information Processing Systems Conference
  (NIPS 2013)}}, Vol.~\bibinfo{volume}{26}. Neural Information Processing
  Systems Foundation (NIPS).
\newblock


\bibitem[\protect\citeauthoryear{Vaswani, Shazeer, Parmar, Uszkoreit, Jones,
  Gomez, Kaiser, and Polosukhin}{Vaswani et~al\mbox{.}}{2017}]%
        {vaswani2017attention}
\bibfield{author}{\bibinfo{person}{Ashish Vaswani}, \bibinfo{person}{Noam
  Shazeer}, \bibinfo{person}{Niki Parmar}, \bibinfo{person}{Jakob Uszkoreit},
  \bibinfo{person}{Llion Jones}, \bibinfo{person}{Aidan~N Gomez},
  \bibinfo{person}{Lukasz Kaiser}, {and} \bibinfo{person}{Illia Polosukhin}.}
  \bibinfo{year}{2017}\natexlab{}.
\newblock \showarticletitle{Attention is All you Need}. In
  \bibinfo{booktitle}{\emph{NIPS}}.
\newblock


\bibitem[\protect\citeauthoryear{Wang, Fu, Fu, and Wang}{Wang
  et~al\mbox{.}}{2017}]%
        {wang2017deep}
\bibfield{author}{\bibinfo{person}{Ruoxi Wang}, \bibinfo{person}{Bin Fu},
  \bibinfo{person}{Gang Fu}, {and} \bibinfo{person}{Mingliang Wang}.}
  \bibinfo{year}{2017}\natexlab{}.
\newblock \showarticletitle{Deep \& cross network for ad click predictions}.
\newblock In \bibinfo{booktitle}{\emph{Proceedings of the ADKDD'17}}.
  \bibinfo{pages}{1--7}.
\newblock


\bibitem[\protect\citeauthoryear{Wei, Wang, Nie, He, and Chua}{Wei
  et~al\mbox{.}}{2020}]%
        {wei2020graph}
\bibfield{author}{\bibinfo{person}{Yinwei Wei}, \bibinfo{person}{Xiang Wang},
  \bibinfo{person}{Liqiang Nie}, \bibinfo{person}{Xiangnan He}, {and}
  \bibinfo{person}{Tat-Seng Chua}.} \bibinfo{year}{2020}\natexlab{}.
\newblock \showarticletitle{Graph-Refined Convolutional Network for Multimedia
  Recommendation with Implicit Feedback}. In
  \bibinfo{booktitle}{\emph{Proceedings of the 28th ACM International
  Conference on Multimedia}}. \bibinfo{pages}{3541--3549}.
\newblock


\bibitem[\protect\citeauthoryear{Xie, Ling, Wang, Wang, Xia, and Lin}{Xie
  et~al\mbox{.}}{2020}]%
        {xie2020deep}
\bibfield{author}{\bibinfo{person}{Ruobing Xie}, \bibinfo{person}{Cheng Ling},
  \bibinfo{person}{Yalong Wang}, \bibinfo{person}{Rui Wang},
  \bibinfo{person}{Feng Xia}, {and} \bibinfo{person}{Leyu Lin}.}
  \bibinfo{year}{2020}\natexlab{}.
\newblock \showarticletitle{Deep Feedback Network for Recommendation}.
\newblock \bibinfo{journal}{\emph{Proceedings of IJCAI-PRICAI}}
  (\bibinfo{year}{2020}).
\newblock


\bibitem[\protect\citeauthoryear{Yu, Ren, Sun, Gu, Sturt, Khandelwal, Norick,
  and Han}{Yu et~al\mbox{.}}{2014}]%
        {yu2014personalized}
\bibfield{author}{\bibinfo{person}{Xiao Yu}, \bibinfo{person}{Xiang Ren},
  \bibinfo{person}{Yizhou Sun}, \bibinfo{person}{Quanquan Gu},
  \bibinfo{person}{Bradley Sturt}, \bibinfo{person}{Urvashi Khandelwal},
  \bibinfo{person}{Brandon Norick}, {and} \bibinfo{person}{Jiawei Han}.}
  \bibinfo{year}{2014}\natexlab{}.
\newblock \showarticletitle{Personalized entity recommendation: A heterogeneous
  information network approach}. In \bibinfo{booktitle}{\emph{Proceedings of
  the 7th ACM international conference on Web search and data mining}}.
  \bibinfo{pages}{283--292}.
\newblock


\bibitem[\protect\citeauthoryear{Zhang, Cao, Zhu, Li, and Sun}{Zhang
  et~al\mbox{.}}{2018}]%
        {zhang2018coupledcf}
\bibfield{author}{\bibinfo{person}{Quangui Zhang}, \bibinfo{person}{Longbing
  Cao}, \bibinfo{person}{Chengzhang Zhu}, \bibinfo{person}{Zhiqiang Li}, {and}
  \bibinfo{person}{Jinguang Sun}.} \bibinfo{year}{2018}\natexlab{}.
\newblock \showarticletitle{Coupledcf: Learning explicit and implicit user-item
  couplings in recommendation for deep collaborative filtering}. In
  \bibinfo{booktitle}{\emph{IJCAI International Joint Conference on Artificial
  Intelligence}}.
\newblock


\bibitem[\protect\citeauthoryear{Zhao, Zhang, Ding, Xia, Tang, and Yin}{Zhao
  et~al\mbox{.}}{2018}]%
        {zhao2018recommendations}
\bibfield{author}{\bibinfo{person}{Xiangyu Zhao}, \bibinfo{person}{Liang
  Zhang}, \bibinfo{person}{Zhuoye Ding}, \bibinfo{person}{Long Xia},
  \bibinfo{person}{Jiliang Tang}, {and} \bibinfo{person}{Dawei Yin}.}
  \bibinfo{year}{2018}\natexlab{}.
\newblock \showarticletitle{Recommendations with negative feedback via pairwise
  deep reinforcement learning}. In \bibinfo{booktitle}{\emph{Proceedings of the
  24th ACM SIGKDD International Conference on Knowledge Discovery \& Data
  Mining}}. \bibinfo{pages}{1040--1048}.
\newblock


\bibitem[\protect\citeauthoryear{Zhou, Bai, Song, Liu, Zhao, Chen, and
  Gao}{Zhou et~al\mbox{.}}{2018a}]%
        {zhou2018atrank}
\bibfield{author}{\bibinfo{person}{Chang Zhou}, \bibinfo{person}{Jinze Bai},
  \bibinfo{person}{Junshuai Song}, \bibinfo{person}{Xiaofei Liu},
  \bibinfo{person}{Zhengchao Zhao}, \bibinfo{person}{Xiusi Chen}, {and}
  \bibinfo{person}{Jun Gao}.} \bibinfo{year}{2018}\natexlab{a}.
\newblock \showarticletitle{Atrank: An attention-based user behavior modeling
  framework for recommendation}. In \bibinfo{booktitle}{\emph{Proceedings of
  the AAAI Conference on Artificial Intelligence}}, Vol.~\bibinfo{volume}{32}.
\newblock


\bibitem[\protect\citeauthoryear{Zhou, Mou, Fan, Pi, Bian, Zhou, Zhu, and
  Gai}{Zhou et~al\mbox{.}}{2019}]%
        {zhou2019deep}
\bibfield{author}{\bibinfo{person}{Guorui Zhou}, \bibinfo{person}{Na Mou},
  \bibinfo{person}{Ying Fan}, \bibinfo{person}{Qi Pi}, \bibinfo{person}{Weijie
  Bian}, \bibinfo{person}{Chang Zhou}, \bibinfo{person}{Xiaoqiang Zhu}, {and}
  \bibinfo{person}{Kun Gai}.} \bibinfo{year}{2019}\natexlab{}.
\newblock \showarticletitle{Deep interest evolution network for click-through
  rate prediction}. In \bibinfo{booktitle}{\emph{Proceedings of the AAAI
  conference on artificial intelligence}}, Vol.~\bibinfo{volume}{33}.
  \bibinfo{pages}{5941--5948}.
\newblock


\bibitem[\protect\citeauthoryear{Zhou, Zhu, Song, Fan, Zhu, Ma, Yan, Jin, Li,
  and Gai}{Zhou et~al\mbox{.}}{2018b}]%
        {zhou2018deep}
\bibfield{author}{\bibinfo{person}{Guorui Zhou}, \bibinfo{person}{Xiaoqiang
  Zhu}, \bibinfo{person}{Chenru Song}, \bibinfo{person}{Ying Fan},
  \bibinfo{person}{Han Zhu}, \bibinfo{person}{Xiao Ma},
  \bibinfo{person}{Yanghui Yan}, \bibinfo{person}{Junqi Jin},
  \bibinfo{person}{Han Li}, {and} \bibinfo{person}{Kun Gai}.}
  \bibinfo{year}{2018}\natexlab{b}.
\newblock \showarticletitle{Deep interest network for click-through rate
  prediction}. In \bibinfo{booktitle}{\emph{Proceedings of the 24th ACM SIGKDD
  International Conference on Knowledge Discovery \& Data Mining}}.
  \bibinfo{pages}{1059--1068}.
\newblock


\end{thebibliography}


\end{document}